\documentclass[conference]{IEEEtran}

\usepackage[dvipsnames]{xcolor}
\usepackage{soul}
\usepackage{xspace}
\usepackage{bm}
\usepackage{stmaryrd}
\usepackage{booktabs}
\usepackage{multirow}
\usepackage{subfigure}
\usepackage{enumitem}
\usepackage{graphicx}
\usepackage{amsmath}
\usepackage{array}
\usepackage{hyperref}
\newcolumntype{C}[1]{>{\centering\arraybackslash}p{#1}}

\usepackage{tikz}
\usepackage{amssymb} 
\newcommand*\circled[1]{\tikz[baseline=(char.base)]{
            \node[shape=circle,draw,inner sep=0.1pt] (char) {#1};}}
\newcommand*\colorcircled[3]{\tikz[baseline=(char.base)]{
    \node[shape=circle,draw,inner sep=0.1pt, fill=#1, text=#2] (char) {#3};}}

\newcommand{\ouralg}{SSNet\xspace}

\newcommand*\emptydot[1][0.7ex]{\tikz\draw[thick] (0,0) circle (#1);} 
\newcommand*\semidot[1][0.7ex]{%
  \begin{tikzpicture}
  \draw[fill] (0,0)-- (90:#1) arc (90:270:#1) -- cycle ;
  \draw[thick] (0,0) circle (#1);
  \end{tikzpicture}}
\newcommand*\soliddot[1][0.8ex]{\tikz\fill (0,0) circle (#1);} 

\begin{document}

\title{\Huge{\ouralg: A Lightweight Multi-Party Computation Scheme for Practical Privacy-Preserving Machine Learning Service in the Cloud}}

\author{\IEEEauthorblockN{Shijin Duan\IEEEauthorrefmark{1},
Chenghong Wang\IEEEauthorrefmark{2},
Hongwu Peng\IEEEauthorrefmark{3}, 
Yukui Luo\IEEEauthorrefmark{4},
Wujie Wen\IEEEauthorrefmark{5},
Caiwen Ding\IEEEauthorrefmark{3},
Xiaolin Xu\IEEEauthorrefmark{1}}
\IEEEauthorblockA{\IEEEauthorrefmark{1}Northeastern University, \IEEEauthorrefmark{2}Indiana University Bloomington, \IEEEauthorrefmark{3}University of Connecticut, }
\IEEEauthorblockA{\IEEEauthorrefmark{4}University of Massachusetts Dartmouth, \IEEEauthorrefmark{5}North Carolina State University}
\IEEEauthorblockA{duan.s@northeastern.edu, cw166@iu.edu, hongwu.peng@uconn.edu, yluo2@umassd.edu, }
\IEEEauthorblockA{wwen2@ncsu.edu, caiwen.ding@uconn.edu, x.xu@northeastern.edu}}

\maketitle

\begin{abstract}
As privacy-preserving becomes a pivotal aspect of deep learning development, multi-party computation (MPC) has gained prominence for its efficiency and strong security. However, the practice of current MPC frameworks is limited, especially when dealing with large neural networks, exemplified by the prolonged execution time of 25.8 seconds for secure inference on ResNet-152. The primary challenge lies in the reliance of current MPC approaches on additive secret sharing, which incurs significant communication overhead with non-linear operations such as comparisons. Furthermore, additive sharing suffers from poor scalability on party size. In contrast, the evolving landscape of MPC necessitates accommodating a larger number of compute parties and ensuring robust performance against malicious activities or computational failures.

In light of these challenges, we propose \ouralg, \ul{which for the first time, employs Shamir's secret sharing (SSS) as the backbone of MPC-based machine learning framework}. We meticulously develop all framework primitives and operations for secure deep-learning models that are tailored to seamlessly integrate with the SSS scheme. \ouralg demonstrates the ability to scale up party numbers straightforwardly and embeds strategies to authenticate the computation correctness without incurring significant performance overhead.  Additionally, \ouralg introduces masking strategies designed to reduce communication overhead associated with non-linear operations. We conduct comprehensive experimental evaluations on commercial cloud computing infrastructure from Amazon AWS, as well as across diverse prevalent DNN model architectures and datasets. \ouralg demonstrates a substantial performance boost, achieving speed-ups ranging from $3\times$ to $14\times$ compared to state-of-the-art MPC frameworks, e.g., secure inference on ResNet-152 consumes just 4.5 seconds. Moreover, \ouralg also represents the first framework that is evaluated on a five-party computation setup, in the context of secure deep learning inference. 
\end{abstract}

\section{Introduction}
The remarkable advancements in deep learning have spurred the rise of cloud-based intelligence, such as Machine Learning-as-a-Service (MLaaS) \cite{weng2022mlaas}. In this emerging business model, the deep neural network (DNN) model vendors deploy their meticulously trained models on cloud platforms, offering services to end users. 
However, such practice raises critical privacy concerns, as user data and model parameters are often valuable and sensitive. 
Ensuring the confidentiality of this information on shared cloud platforms is becoming increasingly important for both users and model vendors.
There are various approaches for secure computing during model inference, including differential privacy (DP) \cite{li2021survey}, homomorphic encryption (HE) \cite{joshi2022comparative}, and multi-party computation (MPC) \cite{zhao2019secure}. Among these methods, MPC is gaining traction thanks to its computational efficiency compared to HE and its stronger security assurances compared to DP. In light of these practical application benefits \cite{tan2021cryptgpu, knott2021crypten, mishra2020delphi}, this work also focuses on MPC-based strategies to enable practical privacy-preserving ML in the cloud.

Recent implementations of MPC-based secure deep learning methods \cite{tan2021cryptgpu, knott2021crypten, wagh2020falcon, kumar2020cryptflow} have primarily adopted additive secret sharing, in which a secret is divided into multiple shares. As a result, the original secret can only be reconstructed by summing these shares. Following this paradigm, the input user data and model parameters in MLaaS are split into multiple parts to specified compute parties. This ensures that no sensitive information on the user data and model is compromised unless a sufficient number of parties collude. During inference, each party independently conducts feed-forward computations of a DNN model. Communication between parties only happens if necessary.
However, the communication between different parties poses a primary challenge in developing modern DNNs in an MPC setup, primarily due to the incurred performance overhead. This challenge is further exacerbated when these multiple parties are distributed across different cloud platforms, rather than being co-located on a single server. Even in a local area network, state-of-the-art (SOTA) MPC solutions, such as \cite{tan2021cryptgpu}, take approximately 26 seconds to process a single image on the MPC-based ResNet-152 model, greatly limiting their practical applicability.
Additionally, there are several other non-quantitative limitations inherent in previous MPC implementations based on additive sharing. We summarize these deficiencies as follows:

\textbf{Efficiency.} Additive sharing schemes in MPC rely on techniques like Beaver triples \cite{beaver1992efficient} to handle \textit{bilinear} operations such as convolution. This inevitably introduces additional computational and communicational burdens. Even variants like replicated secret sharing~\cite{araki2016high}, which allow bilinear operations to be performed locally with minimal communication, present their own challenge, i.e., each party must manage double shares, leading to increased computational complexity. For non-linear operations, such as ReLU, protocols like ABY$^3$ are often employed. These protocols exacerbate the computational and communication overheads. Procedures such as bit extraction and bit injection (for sign-bit extraction and bit-to-arithmetic conversion) are crucial for the implementation of ReLU \cite{tan2021cryptgpu}. They have heavy communication requirements, both in terms of rounds and volume (as detailed in Sec.~\ref{sec:sss_nonlinear} and Tab.~\ref{tab:commcomplexity}). Thus, there is a pressing need for a secure and more efficient protocol to handle non-linear operations on MPC frameworks.

\textbf{Pooling support.} Some previous approaches, e.g., \cite{tan2021cryptgpu}, consider only applying average pooling (AvgPooling) for downsampling in DNNs. This limitation stems from the computational and communication complexities associated with MaxPooling, which requires multiple secret comparisons. These comparisons, as aforementioned in the \textbf{efficiency} issue, are notably time-intensive. Thus, it is desired for another protocol that can efficiently facilitate secure comparisons. Such a protocol should enable both MaxPooling and AvgPooling in GPU-accelerated MPC environments in an efficient manner, broadening the range of downsampling applications in secure DNN inference.

\textbf{Scalability.} Current MPC schemes are typically tailored for a specific number of compute parties, such as 2PC \cite{mishra2020delphi, riazi2019xonn, mohassel2017secureml}, 3PC \cite{tan2021cryptgpu, knott2021crypten, mohassel2018aby3, wagh2020falcon, kumar2020cryptflow}, and 4PC \cite{chaudhari2019trident, byali2019flash}. This limitation naturally arises from secure computation in existing frameworks, which require protocols such as Beaver triples, Garbled Circuits \cite{yao1986generate} and Oblivious Transfer \cite{liu2017oblivious} that are evaluated on a predefined number of parties.
However, the involvement and extensibility of more parties are essential, to provide more flexible computing arrangements and to safeguard against computing failures and malicious parties. An adaptable MPC protocol is necessary, that is inherently scalable, flexible, and configurable to various scenarios and party numbers, for modern and practical MPC environments. {In the event that any party is suspected as invalid or malicious, the protocol ensures the availability of redundant, ready-to-employ alternative parties.}

\textbf{Robustness.} 
In scenarios with a small number of parties, particularly in additive secret sharing, a single party's computing failure or tampering can irreparably compromise secret reconstruction. Assuming a general \textit{n-out-of-n} scheme, additive sharing still requires the correctness in all $n$ parties' computations, offering no fault tolerance. Thus, a more secure and fault-tolerant protocol is essential to address computing failures and potential malicious activities.

Towards comprehensively addressing these concerns, we put forward Shamir's secret sharing (SSS) as a solid alternative for current additive-sharing-based MPC frameworks. Rather than addition, SSS utilizes polynomials to encrypt data. In an $n$PC additive sharing scheme, $n$ shares are required to recover the secret; yet for the $(k,n)$-SSS scheme where $k\leq n$, the secret can be reconstructed by collecting arbitrary $k$ shares, while all $n$ shares vary. This special property bestows flexibility and scalability of the SSS scheme, so that the scalability and robustness issues can be mitigated. Regarding robustness, 
there are approaches to verify the computation correctness \cite{furukawa2019two} and reconstruction \cite{stadler1996publicly} against malicious parties on SSS-based MPC protocol.
Besides, we propose new protocols for the secure comparison in SSS-based DNN, which is sophisticated in previous additive-sharing-based MPC frameworks. Therefore, the efficiency of MPC-based DNN is significantly improved, and naturally, we provide the MaxPooling support, which has been avoided before due to the heavy comparison operations. 

\begin{table}[t]
\centering
\caption{Comparison between \ouralg and other recently-proposed MPC-based deep learning frameworks. ADD means additive secret sharing. \ouralg is the first scheme utilizing SSS as the MPC backbone. $n$PC means \ouralg can be scaled up to an arbitrary number of compute parties $(n\geq 3)$.}
\resizebox{\linewidth}{!}{
\begin{tabular}{l|c|c|cccc|ccc}
\toprule[.5mm]
\multicolumn{1}{c|}{\multirow{2}[4]{*}[-.6cm]{Framework}} & \multirow{2}[4]{*}[-0.9cm]{\rotatebox{90}{Backbone}} & \multirow{2}[4]{*}[-0.7cm]{\rotatebox{90}{\# of Parties}} & \multicolumn{4}{c|}{Operation} & \multicolumn{3}{c}{Evaluation} \\ 
 & & & \rotatebox{90}{Linear/Conv} & \rotatebox{90}{ReLU} & \rotatebox{90}{AvgPooling} & \rotatebox{90}{MaxPooling} & \rotatebox{90}{MNIST/Cifar} & \rotatebox{90}{Tiny ImageNet} & \rotatebox{90}{ImageNet} \\ \midrule
XONN~\cite{riazi2019xonn} & ADD & 2PC & \soliddot & \soliddot & \emptydot & \soliddot & \soliddot & \emptydot & \emptydot \\
Delphi~\cite{mishra2020delphi} & ADD & 2PC & \soliddot & \soliddot & \semidot & \soliddot & \semidot & \emptydot & \emptydot \\
ABY$^3$~\cite{mohassel2018aby3} & ADD & 3PC & \soliddot & \soliddot & \soliddot & \soliddot & \semidot & \emptydot & \emptydot \\
Falcon~\cite{wagh2020falcon} & ADD & 3PC &  \soliddot & \soliddot & \semidot & \soliddot & \soliddot & \semidot & \emptydot \\
CrypTFlow~\cite{kumar2020cryptflow} & ADD & 3PC & \soliddot & \soliddot & \soliddot & \soliddot & \emptydot & \emptydot & \soliddot \\
CryptGPU~\cite{tan2021cryptgpu} & ADD & 3PC & \soliddot & \soliddot & \soliddot & \emptydot & \soliddot & \soliddot & \soliddot \\
\textbf{\ouralg} & \textbf{SSS} & $\bm{n}$\textbf{PC} & \soliddot & \soliddot & \soliddot & \soliddot & \soliddot & \soliddot & \soliddot \\ \bottomrule[.5mm]
\multicolumn{10}{l}{\soliddot\ -- fully supported; \semidot\ -- partially supported or not explicitly explored in}\\
\multicolumn{10}{l}{the work; \emptydot\ -- not supported.}
\end{tabular}}
\label{tab:overview}
\end{table}

{In summary, this work presents the first-ever Shamir's secret sharing-based privacy-preserving machine learning framework, namely \textit{\ouralg}, for efficient MPC model inference in cloud computing.} \ouralg not only significantly mitigates the communication issue in previous MPC frameworks to maximize efficiency, but also demonstrates a linear complexity (per party) when scaling up to an arbitrar number of parties. Similar to the only two known GPU-accelerated MPC frameworks, CrypTFlow~\cite{kumar2020cryptflow} and CryptGPU~\cite{tan2021cryptgpu}, we develop \ouralg for GPU acceleration as well. Tab.~\ref{tab:overview} compares \ouralg and other related frameworks. We highlight the three-fold contributions of this work:

\begin{itemize}[leftmargin=11pt]
    \item \textbf{First-of-its-kind solution.} For the first time, we propose the embodiment using SSS as the backbone of MPC-based deep learning frameworks, our solution shows great robustness and scalability on MPC employment.
    
    \item \textbf{End-to-end design flow.} We showcase the SSS primitives in \ouralg framework, encompassing all essential meta-units for both computation and communication for secure model inference. From the implementation perspective,
    we propose diverse NN modules adaptive to the SSS scheme, which includes the linear operation, the truncation operation, and non-linear operations (e.g., ReLU, MaxPooling, and AvgPooling). These protocols are integrated to enable the secure functionality of modern DNNs. Furthermore, we provide insightful analysis of \ouralg to illustrate its practical design considerations.
    
    \item \textbf{Comprehensive experimental evaluations.} Our model inference evaluation on \ouralg demonstrates that it can achieve $3 \times $ to $14 \times$ speed-up over previous MPC frameworks, and more than $50\%$ communication reduction. Approaching practice, we further evaluate \ouralg on more compute parties (i.e., five-party computation) and in the wide area network setting.
\end{itemize}

\section{Background}
\subsection{Multi-Party Computation on Neural Network}
While MPC is a ubiquitous paradigm for secure computing under various scenarios, the scope of this work focuses on DNN inference of MLaaS. Specifically, secure MPC enables user data and model parameters (as the secret) to be divided into multiple shares, assigned to different parties for joint computation. Each party performs computation on their assigned shares, and no information on the input data or model parameters is leaked unless data from a sufficient number of parties are aggregated. Regarding prior works, we discuss the MPC strategies on two surfaces: secret sharing schemes and secure protocols.

\textbf{Secret sharing}. Additive secret sharing and Shamir's secret sharing (SSS) \cite{shamir1979share} are the two most common schemes in private information distribution. Hereinbefore, the additive sharing is used for the MPC realization on DNN inference, due to its simplicity and efficiency, compared to SSS. For example, CryptGPU \cite{tan2021cryptgpu} and CrypTFlow \cite{kumar2020cryptflow} enabled MPC acceleration on GPU for PyTorch and TensorFlow following additive secret sharing. Unfortunately, additive secret sharing either \circled{1} has low flexibility as a \textit{n-out-of-n} scheme and does not tolerate party computing failure, e.g., common \textit{3-out-of-3} schemes, or \circled{2} asks each party to carry multiple shares thus burdening the computation and memory usage.

\textbf{Secure protocol.} As the cornerstone of MPC, secure protocols ensure data privacy during collaborative computations. Techniques such as ABY$^3$ \cite{mohassel2018aby3}, Garbled Circuits \cite{yao1986generate, ball2016garbling}, and Oblivious Transfer \cite{liu2017oblivious} facilitate MPC across various applications. For example, GryptGPU and CrypTFlow adopted the ABY-based protocols for secure multiplications in the three-party case. Moreover, Ball \textit{et al.} \cite{ball2019garbled} confirmed the efficacy of garbled circuits for two-party DNN inference. However, ABY-based protocols are communication-intensive, particularly when processing the conversion between bit-operation and arithmetic operations, which impede model inference. Besides, garbled circuits require extensive pre-computation, especially for handling large inputs and intricate functions.

\subsection{Shamir Secret Sharing}\label{sec:sss_intro}
Shamir's secret sharing (SSS) has been proposed to divide a secret $a_0$ into multiple parts, namely shares \cite{shamir1979share}. Specifically, assuming a secret is divided to $n$ shares, SSS defines a $(k,n)$ \textit{threshold scheme} as \circled{1} knowledge of any $k$-out-of-$n$ or more shares can reconstruct the original secret \circled{2} knowledge less than $k$ shares always leave the secret undetermined, $n\geq k$. SSS method has information-theoretic security \cite{diffie2022new}, which is secure against unlimited
computing time and resources. The SSS scheme is based on polynomial interpolation within a finite field $\mathbb{F}_p$ containing $p$ numbers, where $p$ is a large prime. Following this setting, we specify $\mathbb{F}_p$ as the \textbf{\textit{finite integer field}} $[0,p-1]$, to favor the integer computation. By using the finite field $\mathbb{F}_p$ to represent a signed range from $-\frac{p-1}{2}$ to $\frac{p-1}{2}$, e.g., $[-5,5]$ for $p=11$, we simply decode an arbitrary integer $x\in \mathbb{F}_p$ as $x\cdot(x\leq q/2)+(x-p)\cdot(x>q/2)$. Conversely, a signed number is first encoded as a number in $\mathbb{F}_p$ to participate in the SSS computation. 
Compared with additive secret sharing, SSS has its inconvenience lying in the polynomial computation during sharing generation and reconstruction. Nevertheless, this computation overhead can be greatly mitigated by offline preprocessing. Moreover, this preprocessing can be conducted once for a sequence of model inferences, and not performed each time of user's input.

Here we use an illustrative example, to compare the robustness and flexibility of SSS v.s. additive secret sharing, following their respective basic schemes. Considering a $(k,n)$ threshold scheme, SSS will generate shares through polynomial interpolation, so any two-out-of-$n$ shares are different. During secret reconstruction, the result can be correctly derived, as long as the number of corrupted parties is less than $k$ and $n\geq 3k$ \cite{furukawa2019two}, and the reconstruction procedure is verifiable \cite{stadler1996publicly}. Thus, a large $n$ can guarantee the flexibility and robustness of SSS during MPC. On the other hand, additive secret sharing requires the summation of $n$ shares (with modulo on an integer ring) equal to the secret, so any corrupted share will render a wrong reconstruction. 

\section{Overview}
\subsection{Threat Model}
\ouralg is architected upon the classical outsourced MPC model~\cite{kamara2011outsourcing}. Within this paradigm, a set of secret owners (i.e., model vendors and users) securely provision their confidential data to a consortium of non-colluding servers (compute nodes/parties), where both the model vendor and the user do not expect their provisioned secrets (models, inference queries) exposed to others. The compute parties jointly evaluate an MPC protocol, perform secure computations on the outsourced secrets, and disclose the outputs exclusively to specified, authorized users. 
In general, there could exist admissible and \textit{semi-honest}~\cite{mohassel2017secureml} adversary who does not deviate from the defined protocol, such as tampering with computing data. However, the adversary is capable of corrupting all but one of the secret owners (i.e., model vendors and users), and at most $(k-1)$ out of $n$ compute parties. The proposed \ouralg should ensure that such an adversary learns nothing about uncorrupted parties' secrets by observing protocol execution transcripts. {Besides, there is a trusted source (a.k.a. trusted third-party in some work \cite{knott2021crypten}) generating and distributing constant shares, which is not corrupted by adversary.}
These assumptions are popular in MPC-based DNN inference and aligns with most works \cite{tan2021cryptgpu, knott2021crypten, kumar2020cryptflow}, while another scenario \cite{mishra2020delphi} can be reduced from ours by making model parameters unprotected and public.

\subsection{Overview of \ouralg Application}
Under \ouralg framework, the secrets of the DNN model and the user's input are encrypted as Shamir's shares, and distributed to all parties. Specifically, the model vendor and the user follow the $(k,n)$-SSS scheme that their secrets are divided into $n$ shares; and $n$ parties are recruited for the confidential DNN inference so that each party holds only \textbf{one} share. 
We adopt $n\geq 2k-1$ in this work, to satisfy the security requirement as discussed in Sec.~\ref{sec:primitive}; yet during evaluation, we set $n=2k-1$ to minimize the communication during secure inference and ensure security under the \textit{semi-honest} assumption. Specifically, when $(k,n)=(2,3)$, this configuration is reduced to the 3-party computation in previous work, but \ouralg does support a larger number of parties. Although our semi-honest assumption does not involve the robustness advantage of SSS (against malicious party), it aligns with the MPC assumptions in previous work; yet previous research about malicious party is still applicable if adding corresponding protocols to our \ouralg. A discussion on this is provided in Sec.~\ref{sec:malicious_party}.
Please note that as the security issue indeed happens on cloud platforms, such as networking \cite{liu2019survey}, virtualization \cite{pearce2013virtualization}, and hardware \cite{xiao2016one} as ongoing research, we focus on the confidential computing of DNN, which is orthogonal to these directions.

\section{\ouralg: Primitives}\label{sec:primitive}
We depict SSS fundamental computation \cite{ben2019completeness}, as they are the primitives of the \ouralg framework.
Further, we present the specialized SSS degree reduction and re-randomization strategy adopted in \ouralg.

\subsection{SSS Fundamentals}
\subsubsection{Share generation $\mathsf{GEN}(a_0,k,n)$} The $n$ shares of $a_0$ are generated by randomly interpolating a $k-1$ degree polynomial. For the $i$-th share, where $i\in[1,n]$, it is generated from
\begin{equation}
    \llbracket a_0\rrbracket_i = (a_0+a_1\cdot \mathtt{ID}_i+...+a_{k-1}\cdot  \mathtt{ID}_i^{k-1})\ \text{mod}\ p
\label{eq:generation}
\end{equation}
While $a_0$ is the secret, $a_1...a_{k-1}\in \mathbb{F}_p$ are randomly generated and identical in all shares. 
Each share/party has a \textit{unique} ID (indicating the share's identity), which can be any non-zero integer in the finite field and is not related to the security. We assume $\mathtt{ID}_i=i$ in our later evaluation for simplicity. Therefrom, the $i$-th share in Eq.\ref{eq:generation} can also be defined as $\llbracket a_0\rrbracket_i=(a_0+a_1 i+...+a_{k-1}i^{k-1})\ \text{mod}\ p, i\in[1,n]$. In this paper, we generically denote the shares of a secret $a_0$ as $\llbracket a_0\rrbracket$ and the $i$-th share as $\llbracket a_0\rrbracket_i$.

\subsubsection{Secret reconstruction $\mathsf{REC}(\llbracket a_0\rrbracket, k, m)$} To recover the secret $a_0$ with degree $(k-1)$, arbitrary $m(\geq k)$ shares are selected. Usually, we choose $m=k$ to minimize the computation and communication, e.g., the first $k$ shares, $i\in[1,k]$. The reconstruction is based on Lagrange's interpolation ~\cite{grunwald1942theory} for the constant coefficient ($\mathtt{ID}=0$):
\begin{equation}
    a_0 = \left (\sum_{i=1}^k \llbracket a_0\rrbracket_{i} \prod_{j=1, j\neq i}^{k}\frac{\mathtt{ID}_j}{\mathtt{ID}_j-\mathtt{ID}_i}\right ) \text{mod}\ p
\label{eq:reconstruction}
\end{equation}
For each term of summation, the multiplier of $\llbracket a_0\rrbracket_{i}$ is a product only related to share IDs, thus it can be prepared offline as a constant value and shared among all parties.
Please note that all arithmetic in SSS is in the finite field $\mathbb{F}_p$, thus we use ``mod $p$'' to reflect this property. For the finite field arithmetic \cite{lidl1994introduction}, we provide a quick reference here: addition/subtraction/multiplication between two operands $a$ and $b$ are performed as $(a+b/a-b/a\times b)\ \text{mod}\ p$, and the division is expressed as $a\times b^{-1}\ \text{mod}\ p$, where the multiplicative inverse of $b$ is derived from $(b\times b^{-1})\equiv1\ (\text{mod}\ p)$. Out of brevity, we omit the modulo operation term in the following content, but it is indeed involved in all the mentioned computations.

\subsubsection{SSS arithmetic}\label{sec:sss_arithmetic} The arithmetic on two secret shares $\llbracket a_0\rrbracket$ and $\llbracket b_0\rrbracket$ follows both properties of SSS and finite field. Specifically, the homomorphism of encrypted computing is guaranteed automatically within the SSS principle:
\begin{itemize}
\setlength{\itemindent}{2em}
    \item $\mathsf{REC}(\llbracket a_0\rrbracket+\llbracket b_0\rrbracket)=a_0+b_0$
    \item $\mathsf{REC}(\llbracket a_0\rrbracket-\llbracket b_0\rrbracket)=a_0-b_0$
    \item $\mathsf{REC}(\llbracket a_0\rrbracket\times\llbracket b_0\rrbracket)=a_0\times b_0$
    \item $\mathsf{REC}(\llbracket a_0\rrbracket\times\llbracket b_0^{-1}\rrbracket)=a_0\times b_0^{-1}=a_0/b_0$
\end{itemize}
This property is also held if $\llbracket b_0\rrbracket$ is a scalar (not share) in $\mathbb{F}_p$. 
Please note that the division is tried to be avoided in SSS, this is because the reconstruction will be distorted if $a_0$ is not divisible by $b_0$, e.g., $\mathsf{REC}(\llbracket 5\rrbracket\times\llbracket 2^{-1}\rrbracket)=5\times 2^{-1}=8$, in $\mathbb{F}_{11}$. Fortunately, the forward computing of DNNs only encompasses addition and multiplication in most cases.

\subsection{SSS Degree Reduction} 
The mechanism of SSS reconstruction guarantees its security on the uniqueness of the solution to Eq.\ref{eq:generation}, i.e., 
there is only one and correct solution for $a_0$ by solving the system of linear equations composed of $\llbracket a_0\rrbracket_1...\llbracket a_0\rrbracket_k$ since the polynomial is of $k-1$ degree. However, during the \textit{bilinear} multiplication of two shares $\llbracket a_0\rrbracket$ and $\llbracket b_0\rrbracket$, $\llbracket a_0\rrbracket_i\times \llbracket b_0\rrbracket_i$ will increase the degree of  Eq.\ref{eq:generation} polynomial from $k-1$ to $2k-2$. As a result, at least $2k-1$ shares are required to construct the result $c_0=a_0\times b_0$. For cascaded multiplications, such as the computing along neural network layers, the polynomial degree as well as the number of required shares will explode exponentially. Under this consideration, the degree reduction for SSS is necessary to ensure the number of shares on an appropriate scale. For a given secret shares $\llbracket c_0\rrbracket$, we denote this operation as $\mathsf{RED}(\llbracket c_0\rrbracket)$, reducing $\llbracket c_0\rrbracket$ to $(k-1)$-degree polynomial. Regarding some solutions for the degree reduction \cite{ben2019completeness, watanabe2015secrecy, shingu2016secrecy}, we adopt the method from \cite{ben2019completeness} because it has the most lightweight computation and least communications although indeed requiring more shares. 

In Ben-Or's method \cite{ben2019completeness}, $2k-1$ shares are required for the degree reduction, i.e., $n\geq (2k-1)$. We denote the multiplication result as $\llbracket c_0\rrbracket=\llbracket a_0\rrbracket\times \llbracket b_0\rrbracket$, and the degree-reduced output is $\llbracket c_0\rrbracket'=\mathsf{RED}(\llbracket c_0\rrbracket)$ that only requires $k$ shares.
The degree reduction is expressed as
\begin{equation}
    \llbracket c_0\rrbracket'=\llbracket c_0\rrbracket^T\times (B^{-1}PB)
    \label{eq:degree_reduction}
\end{equation}
$B$ is the Vandermonde matrix and $P$ is the projection matrix. To be specific, $B=(b_{i,j})$ where $b_{i,j}=(\mathtt{ID}_j)^{i-1}, i,j\in[1,2k-1]$, and $P=(p_{i,j})$ where $p_{i,i}=1$ for $i\in[1,k]$, and $p_{i,j}=0$ otherwise. We denote this degree-reducing matrix as $R=B^{-1}PB$, which is a constant matrix and can be computed offline for direct use. Note that, unlike SSS arithmetic that each share is only computed locally, \textit{the degree reduction requires $2k-1$ shares collected to perform Eq.\ref{eq:degree_reduction}}. 
Therefore, the degree reduction protocol must be designed carefully; otherwise, anyone who collected the shares is able to reconstruct the secret $c_0$. On the other hand, although the multiplication (or convolution) output $c_0$ will not leak the information about $a_0$ and $b_0$ directly, we still stress the importance of the privacy of $c_0$. An intuitive explanation is provided in Appendix A.

\textbf{Remark: How does the na\"ive method fail?}
Following Eq.\ref{eq:degree_reduction}, the first step is to multiply $\llbracket c_0\rrbracket_i$ with the $i$-th row of $R$, then making the elements of the $j$-th column summed up to derive $\llbracket c_0\rrbracket'_j$. One na\"ive method is to make the first step executed on each party locally, without leaking any information; for the second step, the $j$-th party ($j\in[1,n], j\neq i$) will send the $i$-th element to the $i$-th party, so that the $i$-th party collects all elements for summation. We take the $(2,3)$-scheme as an example, whose degree reduction on $\llbracket c_0\rrbracket$ is expressed as:
\begin{equation*}
\begin{split}
     &\begin{matrix}
\textbf{P1:}\ \llbracket c_0\rrbracket_1 \cdot[R_{1,1}, R_{1,2}, R_{1,3}]\\
\textbf{P2:}\ \llbracket c_0\rrbracket_2 \cdot[R_{2,1}, R_{2,2}, R_{2,3}]\\
\textbf{P3:}\ \llbracket c_0\rrbracket_3 \cdot[R_{3,1}, R_{3,2}, R_{3,3}]
\end{matrix}\xrightarrow[]{Comm.}\\
&\begin{matrix}
\textbf{P1:}\ [\llbracket c_0\rrbracket_1R_{1,1}, \llbracket c_0\rrbracket_2R_{2,1}, \llbracket c_0\rrbracket_3R_{3,1}] \\
\textbf{P2:}\ [\llbracket c_0\rrbracket_1R_{1,2}, \llbracket c_0\rrbracket_2R_{2,2}, \llbracket c_0\rrbracket_3R_{3,2}] \\
\textbf{P3:}\ [\llbracket c_0\rrbracket_1R_{1,3}, \llbracket c_0\rrbracket_2R_{2,3}, \llbracket c_0\rrbracket_3R_{3,3}]
\end{matrix}\xrightarrow[]{\sum}\begin{matrix}
\llbracket c_0\rrbracket'_1 \\
\llbracket c_0\rrbracket'_2 \\
\llbracket c_0\rrbracket'_3
\end{matrix}
\end{split}
\end{equation*}
The vulnerability of this protocol stems from the public accessibility of the reducing matrix $R$. Taking party P1 as an example, with the knowledge of $R_{2,1}$ and $(\llbracket c_0\rrbracket_2R_{2,1}\ \text{mod}\ p)$, $\llbracket c_0\rrbracket_2$ can be determined uniquely and accurately, due to the reversibility of multiplication on $\mathbb{F}_p$. The same applies for $\llbracket c_0\rrbracket_3$. Consequently, P1, along with the other parties, can reconstruct the secret $c_0$, making this na\"ive protocol insecure.

\begin{figure}[tb!]
    \centering
    \includegraphics[width=.9\linewidth]{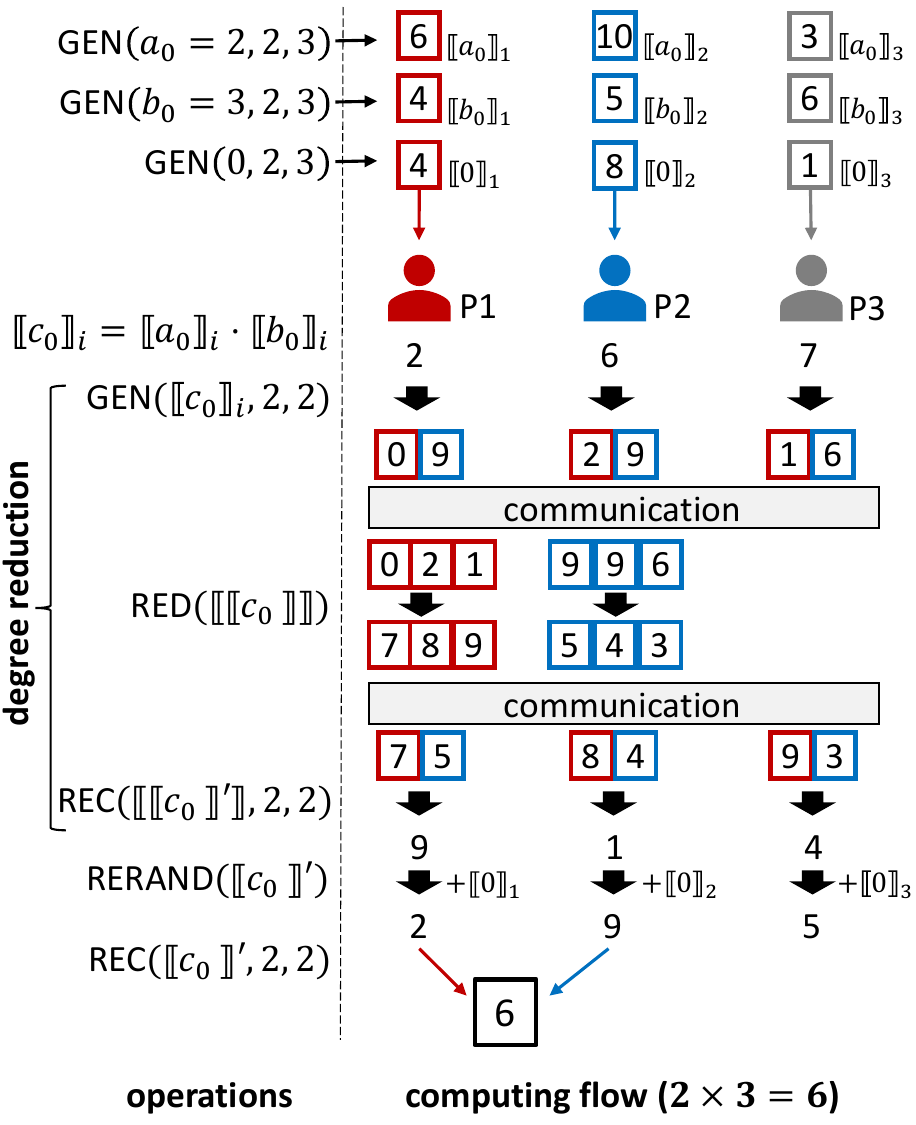}
    \caption{A numerical example for the SSS degree reduction ($\mathsf{RED}$) and re-randomization ($\mathsf{RERAND}$), which computing the multiplication of $a_0=2$ and $b_0=3$. We assume 3 parties in the MPC with $(2,3)$-SSS scheme. Out of brevity, we set the finite field as $\mathbb{F}_{11}$, containing integers $[0,10]$.}
    \label{fig:red_rerand}
\end{figure}

\textbf{Our proposed \textit{reshare} method.}
We propose the following protocol to conduct the degree reduction of $\llbracket c_0\rrbracket$, while  keeping $c_0$ secure (see ``\textbf{degree reduction}'' in Fig.~\ref{fig:red_rerand}):

\begin{enumerate}[leftmargin=11pt]
    \item \textbf{On $n$ parties:} Each party $i$ ($i\in[1,n]$) holds one share of $c_0$, i.e., $\llbracket c_0\rrbracket_i$. All parties treat their share as secret and regenerate Shamir's shares for $k$ selected parties, $\llbracket\llbracket c_0\rrbracket_i\rrbracket=\mathsf{GEN}(\llbracket c_0\rrbracket_i, k, k)$, e.g., P1 to P$k$. Then, the shares $\llbracket\llbracket c_0\rrbracket_i\rrbracket$ in each party $i$ are distributed to the specified $k$ parties.
    
    \item \textbf{On $k$ parties:} 
    After communication, party $i$ in $k$ parties $(i\in[1,k])$ holds shares $\{\llbracket\llbracket c_0\rrbracket_1\rrbracket_i, \llbracket\llbracket c_0\rrbracket_2\rrbracket_i,...,\llbracket\llbracket c_0\rrbracket_n\rrbracket_i\}$. Since the matrix $B^{-1}PB$ is public, each party also has this matrix.
    The $k$ parties perform degree reduction (Eq.\ref{eq:degree_reduction}) with obtained shares, thus will get the SSS format of $\llbracket c_0\rrbracket'$. We denote the outputs each party $i$ is holding as $\{\llbracket\llbracket c_0\rrbracket_1'\rrbracket_i, \llbracket\llbracket c_0\rrbracket_2'\rrbracket_i,...,\llbracket\llbracket c_0\rrbracket_n'\rrbracket_i\}$.
    \item \textbf{On $n$ parties:} The $k$ parties who are holding shares distribute them to all $n$ parties, 
    such that each party $i$ $(i\in[1,n])$ will hold the shares of $\llbracket c_0\rrbracket_i'$, i.e., $\{\llbracket\llbracket c_0\rrbracket_i'\rrbracket_1, \llbracket\llbracket c_0\rrbracket_i'\rrbracket_2,...,\llbracket\llbracket c_0\rrbracket_i'\rrbracket_k\}$. Therefore, party $i$ will execute $\mathsf{REC}(\llbracket\llbracket c_0\rrbracket_i'\rrbracket, k, k)$ to     reconstruct $\llbracket c_0\rrbracket_i'$.
\end{enumerate}
The correctness of this protocol can be proved by the homomorphism of SSS arithmetic. By regarding $\llbracket c_0\rrbracket$ as secrets and generate their shares, this protocol executes $\mathsf{REC}(\llbracket\llbracket c_0\rrbracket\rrbracket^T\times (B^{-1}PB))=\llbracket c_0\rrbracket^T\times (B^{-1}PB)$ in a secured process, i.e., each party only has one share of $\llbracket c_0\rrbracket$.
Consequently, the degree reduction is performed without leaking the knowledge of $\llbracket c_0\rrbracket$.

\subsection{SSS Re-randomization}\label{sec:rerandomization} 
During SSS multiplication, the coefficients of product $\llbracket c_0\rrbracket$ are totally determined by the coefficients of $\llbracket a_0\rrbracket$ and $\llbracket b_0\rrbracket$, e.g., $c_1=(a_0b_1+a_1b_0)$, following the representation of Eq.\ref{eq:generation}. Therefore, the coefficients of $\llbracket c_0\rrbracket'$, i.e., the truncated version of $\llbracket c_0\rrbracket$ after the degree reduction, is also deterministic. To guarantee the information-theoretic security of SSS (see Sec.~\ref{sec:sss_intro}), all coefficients of $\llbracket c_0\rrbracket'$ should be completely random. We adopt the randomization method proposed in \cite{ben2019completeness} under our \ouralg framework. Notably, rather than generating Shamir's shares in each party and involving tedious communication like \cite{ben2019completeness}, we only need one-time SSS generation and distribution, as a pre-processing of model inference. 
Specifically, shares for 0 are generated with $k-1$ degree, i.e., $\llbracket0\rrbracket=\mathsf{GEN}(0, k, n)$, and distributed to each party $i$, who will later add $\llbracket0\rrbracket_i$ onto $\llbracket c_0\rrbracket_i'$. Since the constant term of $\llbracket0\rrbracket$ is zero, $\llbracket c_0\rrbracket' + \llbracket0\rrbracket$ keeps the secret $c_0$ unchanged; yet, the other coefficients are completely random because $\llbracket0\rrbracket$ have random coefficients. Thus, the information security of $\llbracket c_0\rrbracket_i'$ is promised. We denote this protocol as $\mathsf{RERAND}(\llbracket c_0\rrbracket_i')$.

We demonstrate a simple numerical example in Fig.\ref{fig:red_rerand}, in order to visualize the protocols for SSS degree reduction and re-randomization. It performs a completed SSS multiplication under a 3-party scenario with (2,3)-SSS scheme. The degree reduction involves two data exchanges through P2P (peer-to-peer) channels between parties, while re-randomization only requires local computation. Before degree reduction, all the 3 shares (i.e., $2,6,7$) are needed to reconstruct the correct result (i.e., $6$); yet after $\mathsf{RED}(\cdot)$ and $\mathsf{RERAND}(\cdot)$, only the first two shares (i.e., $2,9$) are collected for correct reconstruction.

\section{\ouralg: Communication Layout}
\begin{figure}[tb!]
    \centering
    \includegraphics[width=\linewidth]{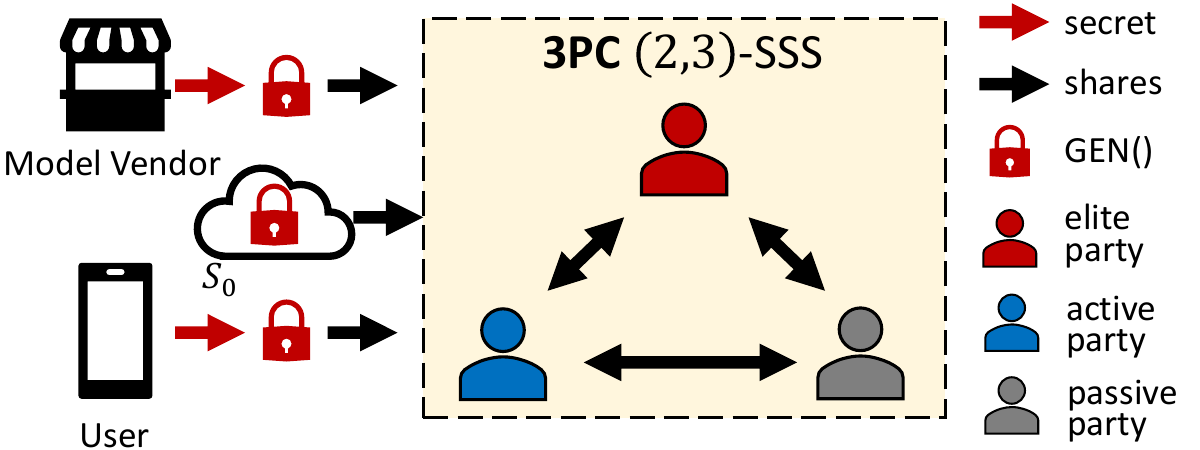}
    \caption{The dataflow for \ouralg framework. }
    \label{fig:dataflow}
\end{figure}

In this section, we discuss the dataflow of \ouralg with a focus on the inter-party interactions. 
Without loss of generality, a 3PC case study employing $(2,3)$-SSS scheme is presented, as shown in Fig~\ref{fig:dataflow}. Specifically, the model vendor provides a general DNN model, including but not limited to models for classification and object detection tasks, and the user uploads the inference query. 
To ensure confidentiality, both the model and the inference query are secret-shared among the computing parties.
Besides, we consider there exists a trusted source\footnote{In \ouralg, we rely on a trusted source to supply these shares for efficiency reasons. However, one can trivially replace this with existing secure protocols~\cite{furukawa2019two, abraham2020blinder}, to achieve a stronger security model.} who supply zero-shares, i.e., $\llbracket 0\rrbracket$, and other auxiliary secret shares to assist the secure computations. By trust, we say that the randomness for generating these shares cannot be obtained, controlled, or tampered with by any external parties. For example, these shares can be derived within a TEE~\cite{yandamuri2021communication}, where secret share randomness is exclusively available inside a hardware isolated region. 

\subsection{Party Management}
We divide the compute parties into three categories, \textit{elite}, \textit{active}, and \textit{passive}. In a general $(k,n)$ scheme, one elite and $k-1$ active parties are selected, while other parties are marked as passive. 
These threes categories are defined as:
\begin{itemize}
    \item \textbf{Elite:} The party who participates in the entire MPC, and performs the heaviest computing load, including local computation, share generation, and secret reconstruction. $(\#=1)$
    \item \textbf{Active:} The party who participates in the entire MPC, but only performs partial computation, without operations such as $\mathsf{GEN}(\cdot)$ and $\mathsf{REC}(\cdot)$. $(\#=k-1)$
    \item \textbf{Passive:} The party who only participates in the SSS multiplication-related computing. $(\#=n-k)$
\end{itemize}
Detailed cooperation policy is discussed in Sec.~\ref{sec:secure_prim_sssnet}. 
Although the elite party performs secret reconstruction, we develop corresponding protections preventing the elite party from knowing the secrets even after reconstruction. 

For the party selection, while even the elite party cannot recover the secrets under our protocol, it is encouraged to rotate the roles of compute parties during the \ouralg framework running. This will balance the computing workload and the security level of all parties. However, other election strategies can be applied depending on different computing budgets. In the following context, we take P1 as elite, P2-P$k$ as active, and the rest parties as passive. While the above discussions are orthogonal to this work, we briefly demonstrate them for subsequent development.

\textbf{Remark: An alternative to $S_0$.} Except our strategy introducing a trusted sever $S_0$, we also provide an alternative only with $n$ compute parties but requiring extra communication, that was commonly adopted in previous SSS protocol~\cite{ben2019completeness}: taking zero shares as an example, to generate shares of $0$ that still follows the $(k,n)$-SSS scheme, each party $i$ generates shares $\llbracket 0\rrbracket^i$, and they exchange shares to each other and perform summation, so that the final share on party $j$ will be $\sum_{i=1}^N\llbracket 0\rrbracket^i_j = \llbracket \sum_{i=1}^N 0\rrbracket^i_j$. Here secret 0 can be a random number for the masking shares (Sec.~\ref{sec:secure_prim_sssnet}). Specifically, each party randomly generates a number following the rule of generation; after exchange and summation, the final shares point to the sum of these random numbers which still falls into the rule of generation. Since the share generation is a preprocessing step, where the communication complexity is not the major consideration, both these two methods are suitable under our \ouralg framework. Nevertheless, our strategy requires much less communication, which significantly alleviates the preparation efforts when the networking environment between parties is of high-latency.

\begin{figure*}[t]
    \centering
\subfigure[SSS-Linear]{\centering
    \includegraphics[width=0.315\linewidth]{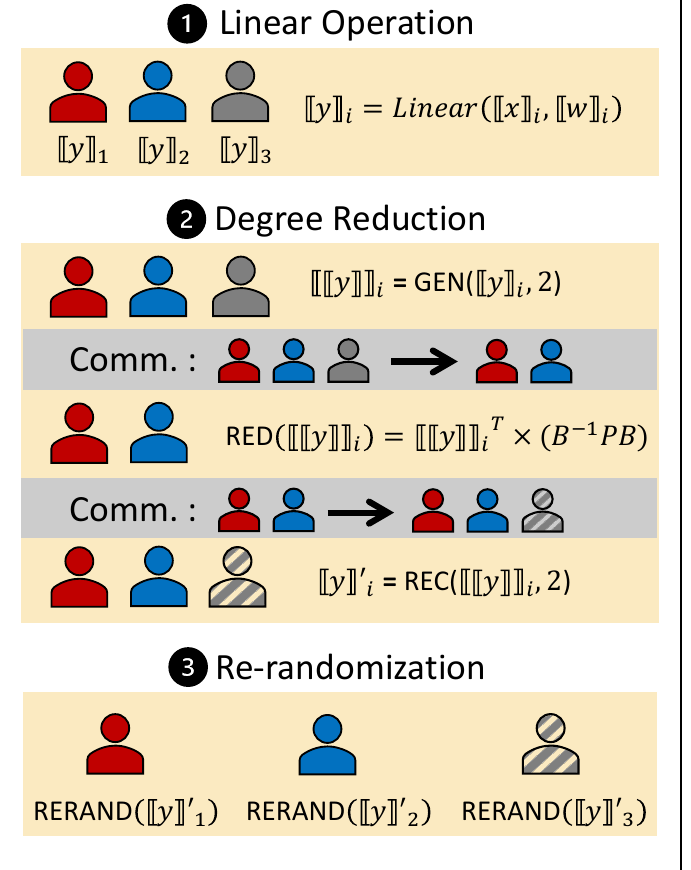}}
\subfigure[SSS-Truncation]{\centering
\includegraphics[width=0.33\linewidth]{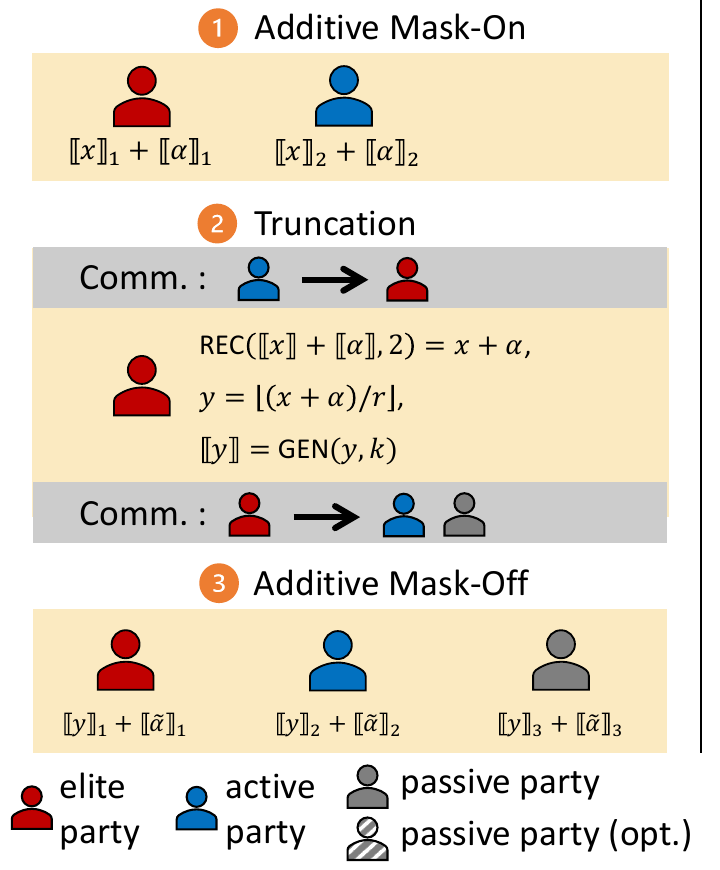}}
\subfigure[SSS-NonLinear]{\centering
    \includegraphics[width=0.315\linewidth]{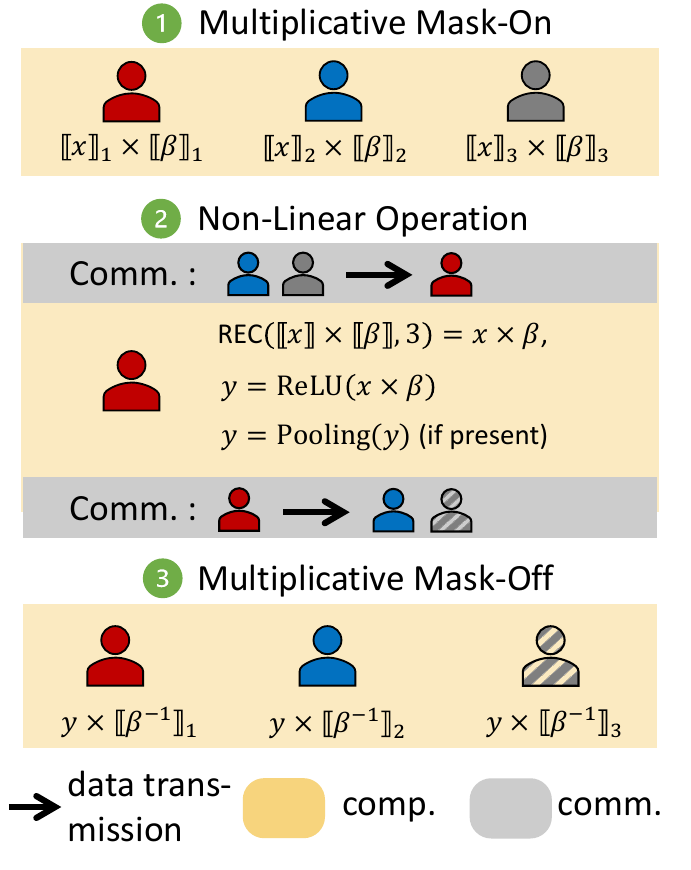}}
    \caption{The SSS operations in our \ouralg framework. We take a $(2,3)$-SSS scheme as an example. Note that the passive party with slash pattern means it is optional, based on the subsequent SSS operation. As denotation, comp. is for computation, and comm. is for communication.}
    \label{fig:SSSOpeartion}
\end{figure*}

\subsection{Malicious Parties}\label{sec:malicious_party}
{Under the $(k,n)$-SSS scheme, \ouralg adopts the minimum requirement $n=2k-1$ to maximize the efficiency. However, with the concern of fault tolerance and malicious parties, more parties should be involved. For example, with the existence of malicious adversary, results from compute parties could be tampered with. Thus, authentication methods on the protocol can be introduced, verifying if the computation results (especially for multiplication) are modified or deviate from the $(k,n)$ scheme. Approaches such as \cite{furukawa2019two} can check the correctness of computation, if no more than one-third of parties are malicious without further overhead than the semi-honest assumption. In this case, more compute parties are required, e.g., $n>3k$, to evaluate the computation again on other legitimate parties.}

\section{\ouralg: Secure Modules}
\label{sec:secure_prim_sssnet}
In this section, we demonstrate the SSS modules of their NN counterpart. \ouralg can be applied to all DNNs that are composed of the following modules. 
Since SSS is employed in the finite field, all computations must be in $\mathbb{F}_p$ as well, which is easy to achieve under modern quantization techniques. 
As a unified notation, we use ${x}$ and ${y}$ to represent the input and output of each operation, thus $\llbracket x\rrbracket$ and $\llbracket y\rrbracket$ are the Shamir's shares. Similarly, $\llbracket w\rrbracket$ represents the shares of weights in the operation, if presented. We summarize the operations in Fig.~\ref{fig:SSSOpeartion} with a $(2,3)$-SSS example.

\subsection{SSS-Linear}
The linear layer in neural networks specifies the linear transformation of the input to a new dimension, e.g., the convolution layer and dense layer. 
The computing stream of linear layers (\colorcircled{black}{white}{1}) shown in Fig.~\ref{fig:SSSOpeartion}(a) can be expressed as $\llbracket y\rrbracket=Dense(\llbracket x\rrbracket, \llbracket w\rrbracket)$ for dense layer and $\llbracket y\rrbracket = Conv2D(\llbracket x\rrbracket, \llbracket w\rrbracket)$ for convolution layer, respectively.
Since \textit{bilinear} SSS multiplication will increase the degree of SSS from $k$ to $2k-1$, degree reduction should be applied so that the SSS degree will not explode through cascaded layers, where two rounds of communications are involved. During degree reduction ($\mathsf{RED}$, \colorcircled{black}{white}{2}), the elite party, active parties, and $k-1$ passive parties participate in computing and communicating, yet the followed re-randomization ($\mathsf{RERAND}$, \colorcircled{black}{white}{3}) is at least performed on the elite party and active parties, since degree-reduced secret $\llbracket y\rrbracket'$ can be recovered by only $k$ parties, as illustrated in Fig.~\ref{fig:red_rerand}. The $\mathsf{RERAND}$ on passive parties is optional, based on the subsequent SSS operation, i.e., NonLinear or Truncation.
Note that the zero shares in re-randomization come from the trusted server $S_0$, which is transferred to compute parties in advance. 

\subsection{SSS-Truncation}
Truncation is necessary after multiplication in a quantized neural network. 
In previous works that employ additive sharing, truncation is conducted straightforwardly and locally \cite{tan2021cryptgpu}. Although it will induce small errors by rounding to an integer, additive sharing has a good tolerance for this, while SSS does not. This is because the secret of SSS could be distorted during division as we discussed in Sec.~\ref{sec:sss_arithmetic}.
We also provide an intuitive example of this in Appendix B.
Therefore, an error-free computing method is necessary for SSS, which is introduced in our work in Fig.~\ref{fig:SSSOpeartion}(b). 

For an input $x$, the truncation operation derives the quantized version of $x$ as $y=\left \lfloor x/r \right \rfloor $ where $r$ is a pre-defined scaling factor; equivalently, we express $x=r\cdot y+b, b\in[0..r)$. In order to perform quantization on secured data, we adopt a masking strategy. Specifically, the trusted server $S_0$ pre-generates two set of masks $\alpha$ and $\widetilde{\alpha}$ with the same size of $x$, under the condition that $\alpha$ must be a multiple of $r$, i.e., $\alpha=e\cdot r$ and $\widetilde{\alpha}=-e$. $S_0$ generates the Shamir's shares for the two masks, i.e., $\llbracket \alpha\rrbracket$ and $\llbracket \widetilde{\alpha}\rrbracket$, and distribute them to all parties, in advance of model inference. Thus, the masks have the property $\llbracket0\rrbracket=\llbracket \alpha\rrbracket + \llbracket \widetilde{\alpha}\cdot r\rrbracket$. \colorcircled{Orange}{white}{1} The input data shares $\llbracket x\rrbracket$ are first added with the mask shares $\llbracket \alpha\rrbracket$, i.e., 
\begin{equation}
    \llbracket x\rrbracket+\llbracket \alpha\rrbracket=\llbracket x + \alpha\rrbracket=\llbracket (y+e)\cdot r + b\rrbracket
\end{equation}
\colorcircled{Orange}{white}{2} Then, shares are collected by the elite party to reconstruct the secret $(y+e)\cdot r + b$, and apply the truncation, with result $\left \lfloor ((y+e)\cdot r + b)/r\right \rfloor=y+e$. The elite party also generates the secret shares $\llbracket y+e\rrbracket$ for distribution, following the $(k,n)$-SSS scheme. 
\colorcircled{Orange}{white}{3} The $\llbracket y+e\rrbracket$ are distributed to active and passive parties. This is because the subsequent operations of truncation, i.e., Linear or NonLinear, both require the passive parties for computing. Consequently, all parties conduct the addition on $\llbracket y+e\rrbracket$ and $\llbracket \widetilde{\alpha}\rrbracket$ to recover the final result $\llbracket y\rrbracket$ in the encrypted (i.e., ciphertext) domain. Since the masking and unmasking operation is based on addition, no degree reduction or re-randomization is required.

\subsection{SSS-NonLinear}\label{sec:sss_nonlinear}
By non-linear operation in SSS, we define it in a coarse-grained scale as the post-processing after a linear layer, such as ReLU, AvgPooling, and MaxPooling. These operations contain non-linear calculations on Shamir's shares. In previous work \cite{tan2021cryptgpu, wagh2020falcon, kumar2020cryptflow}, the comparison was commonly performed by the bit extraction following ABY$^3$ protocol \cite{mohassel2018aby3}. As aforementioned, the ABY3 needs many rounds of communication. Further, this protocol is specially designed for 3PC, where the case of more parties has not been carefully investigated. Hereby, we propose another alternative method for comparison operation, whose communication only increases \textbf{linearly} by party number increment.
Our strategy is to generate a positive mask $\beta$ and its reciprocal $\beta^{-1}$, together with their corresponding shares $\llbracket \beta \rrbracket$ and $\llbracket \beta^{-1} \rrbracket$, which are distributed from the trusted server $S_0$. This mask is multiplied by the input shares $\llbracket x\rrbracket$, and the product will be reconstructed, rather than the ${x}$, so that the secret input is kept confidential. An illustration of SSS-NonLinear is shown in Fig.~\ref{fig:SSSOpeartion}(c).

We depict the execution as follows: \colorcircled{Green}{white}{1} For an input share $\llbracket x\rrbracket$ indicating a secure feature map $x$, the shares of positive mask $\beta$ (i.e., $\beta>0$) with the same size of $x$ are prefetched by all parties, so that Hadamard product is conducted between the input $\llbracket x\rrbracket$ and the random positive mask $\llbracket \beta\rrbracket$. When only ReLU is applied as the non-linear operation, positive is the only restriction for $\beta$. However, if AvgPooling/MaxPooling is also contained, we further add another condition, that the generated $\beta$ in one pooling kernel should be the same in order to not affect the followed $\beta^{-1}$ cancelling. For example, if the kernel size is $2\times 2$, then $\beta$ in every $2\times 2$ square block (4 elements) have the same positive number. \colorcircled{Green}{white}{2} The masked input $\llbracket x\rrbracket\times \llbracket \beta\rrbracket$ will be collected by the elite party from active parties and passive parties to reconstruct the $x\times \beta$. Although bilinear SSS multiplication is involved during the Hadamard product, we do not need degree reduction and re-randomization because we have enough parties (in total $2k-1$) and we only need the masked secret $x\times \beta$ not its SSS format for followed computing. The non-linear operations then are performed in plaintext, $y=NonLinear(x\times \beta)$. The confidentiality of $x$ is ensured by $\beta$. \colorcircled{Green}{white}{3} The shares of the reciprocal mask $\llbracket \beta^{-1}\rrbracket$ are also pre-distributed by trusted server $S_0$ to all parties. The output of the non-linear operations is distributed by the elite party to other parties in \textit{plain text}, and the output will be multiplied by the inversed-mask shares $\llbracket y\rrbracket=y \times \llbracket \beta^{-1}\rrbracket$. Since the mask-off operation is conducted by scalar-share multiplication, the polynomial degree is not increased so that no degree reduction is required for post-processing. Based on the subsequent SSS operation, i.e., Linear or Truncation, this step on passive parties is optional. By implementing a positive mask for the non-linear operation and involving encrypted plain-text calculation, there are no complicated communications between parties, which is a great superiority against previous works. Also, our strategy only requires one-round communication, which is significantly more efficient than previous methods \cite{mohassel2018aby3}.

\textbf{Extra step for AvgPooling operation.} While the summation and division are conducted in one kernel during AvgPooling, the division on $x\times \beta$ will greatly limit the selection of $\beta$. Thus, we move the division to the SSS-Truncation operation as a subsequent computation. Assume AvgPooling has kernel size $k_h\times k_w$, we add one more rounding division $y=\left [ y/k_h/k_w \right ]$ after the truncation $y=\left \lfloor (x+\alpha)/r\right \rfloor$. Consequently, only summation is applied in the SSS-NonLinear operation for AvgPooling. Note that this manipulation might incur a small error on the final AvgPooling result if we execute the rounding computation before the summation, e.g., at most $\pm2$ for the $2\times2$ kernel. For a high bit-width network, e.g., 16-bit, the small deviation on the least significant bits is negligible for the model inference performance. Besides, random masks for truncation should be tuned as $\alpha=e\cdot r\cdot k_h\cdot k_w$ accordingly (as multiplies of $k_h\cdot k_w$), to ensure the $\widetilde{\alpha}$ canceling is correctly performed. A detailed analysis is provided in Appendix C.

\begin{table}[t]
\caption{The communication complexity of \ouralg and other related frameworks, in the 3PC case. $\ell$ is the bit size of secret data. We also provide the generalized complexity analysis of \ouralg under $n$PC scenario, where $n=2k-1$ for efficiency under the semi-honest assumption. 
For the communication of \ouralg, ``$\mathbf{A}/\mathbf{B}$'' means if the final Comm. of each operation includes passive parties ($\mathbf{B}$) or not ($\mathbf{A}$), as demonstrated in Fig.~\ref{fig:SSSOpeartion}.
}
\resizebox{\linewidth}{!}{
{
\setlength{\tabcolsep}{2pt} 
\begin{tabular}{cl|cc|cc|cc}
\toprule[.5mm]
\multicolumn{2}{c|}{\multirow{2}{*}{Framework}} & \multicolumn{2}{c|}{Linear} & \multicolumn{2}{c|}{Truncation} & \multicolumn{2}{c}{NonLinear} \\
 &  & Round & Comm. & Round & Comm. & Round & Comm. \\\midrule
\multicolumn{1}{c|}{\multirow{4}{*}{\rotatebox{90}{3PC}}} & BLAZE\cite{patra2020blaze} & $1$ & $3\ell$ & $-$ & $-$ & $4$ & $47\ell$ \\
\multicolumn{1}{c|}{} & Falcon\cite{wagh2020falcon} & $1$ & $4\ell$ & $1$ & $-$ & $5+\text{log}\ell$ & $32\ell$ \\
\multicolumn{1}{c|}{} & CryptGPU\cite{tan2021cryptgpu} & $1$ & $3\ell$ & $2$ & $-$ & $3+\text{log}\ell$ & $-$ \\
\multicolumn{1}{c|}{} & \ouralg & $2$ & $6\ell/8\ell$ & $1$ & $3\ell$ & $1$ & $3\ell/4\ell$ \\ \hline
\multicolumn{1}{c|}{\multirow{1}{*}[0.05cm]{\rotatebox{90}{$n$PC}}} & \ouralg & $2$ & \begin{tabular}{c} $(3k^2-3k)\ell/$\\ $(4k^2-4k)\ell$\end{tabular} & $1$ & $(3k-3)\ell$ & $1$ & \begin{tabular}{c} $(3k-3)\ell/$\\ $(4k-4)\ell$\end{tabular} \\ \bottomrule[.5mm]
\end{tabular}}}
\label{tab:commcomplexity}
\end{table}

\subsection{Theoretical Complexity Analysis}
As the communication takes a significant portion during model inference (see Sec.~\ref{sec:ablation}), we hereby analyze the communication complexity of \ouralg and other related works.
We assume the data has bit size $\ell$, and demonstrate the theoretical communication complexity in Tab. \ref{tab:commcomplexity}. Here the communication is defined as the total data exchange on all parties. Under the 3PC scenario (i.e., $(2,3)$-SSS scheme for \ouralg), the difference between our \ouralg framework and other works that are based on additive sharing and ABY$^3$ protocol is straightforward. On the linear layer computing, additive sharing only requires one round of communication to conform the 2-out-of-3 duplicated sharing \cite{tan2021cryptgpu}; on the other hand, Shamir's sharing requires the extra round for degree reduction, thus doubling the communication compared with additive sharing. However, on the non-linear operations, \ouralg only requires one round\footnote{Round is defined as that one party sends and receives data once.} of data exchange to perform the non-linear operation under masking. Other works based on ABY$^3$ protocol require significantly more rounds to do bit decomposition and bit injection~\cite{mohassel2018aby3}. Moreover, the bit manipulation yields large communication, making the previous non-linear protocols time-consuming, while our strategy is lightweight by applying only positive masking. For the truncation protocol, our masking protocol has similar rounds of communication with the ABY$^3$ protocol. While previous works did not explicitly discuss the theoretical communication, our truncation method indeed involves a low communication. In addition, \ouralg can be straightforwardly extended to MPC with more parties, and we provide the theoretical analysis on a general $n$PC case in Tab.~\ref{tab:commcomplexity}.
The round counts for all operations stay the same as the MPC scale increases. For the communication volume, the total communication of SSS-Linear quadratically increases along the party population, yet the total communication of SSS-NonLinear and SSS-Truncation follows a linear increment. Thus, each party will carry out $O(k)$ communication on SSS-Linear but always conduct $O(1)$ communication on SSS-NonLinear and SSS-Truncation, along with the number of parties increasing. Since previous works are dedicated to 3PC running, there is no reference to compare with for a larger scale, e.g., 5PC (under $(3,5)$-SSS scheme). Still, the fixed rounds and linearly-complicated communication on each party ensures \ouralg to be efficiently scaled to arbitrary parties' computing setup.

\begin{figure}[t]
    \centering
    \includegraphics[width=.8\linewidth]{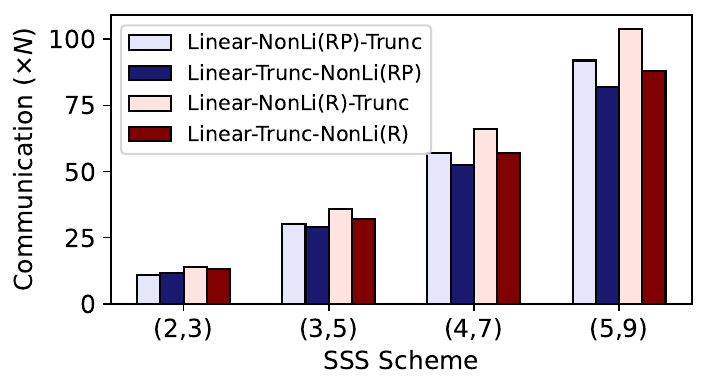}
    \caption{The communication under different operation arrangements and SSS schemes, assuming the data size is $N$. We use RP (ReLU+Pooling) and R (ReLU) to represent the two kinds of NonLi (NonLinear) operations. For Pooling, assume the kernel size is $2\times2$.}
    \label{fig:communication}
\end{figure}

\subsection{Operation Arrangement Analysis}
For a completed DNN layer, where the linear operation is followed by the ReLU function and pooling layer, there are two arrangements for the execution order: Linear-NonLinear-Truncation or Linear-Truncation-NonLinear. Considering the two non-linear cases (ReLU and ReLU+Pooling), we investigate the communication versus the arrangement in Fig.~\ref{fig:communication}. Although the communication round is fixed, the volume of communication is significantly affected by the operations arrangements and the SSS scheme scale. If only ReLU is involved in the SSS-NonLinear operation, it is always better to perform the SSS-Truncation ahead of the SSS-NonLinear. This is because the SSS-Linear undertakes most of the communication; since the input of truncation only requires the elite party and active parties, the Linear-Truncation-ReLU order can reduce communication on SSS-Linear. However, if both ReLU and Pooling are applied to the SSS-NonLinear operation, there is a trade-off in communication. On the one hand, Linear-Truncation pattern can reduce the communication on SSS-Linear; on the other, the data size after Pooling is greatly reduced as well, NonLinear-Truncation pattern makes the communication on both SSS-NonLinear and SSS-Truncation lightweight. 
From Fig.~\ref{fig:communication}, we can observe that for the $(2,3)$-SSS scheme, the Linear-NonLinear-Truncation is a better choice (though the advantage is trivial), while the Linear-Truncation-NonLinear is notably better under other schemes. 

\section{\ouralg: Practical Design Strategies}

\subsection{Bit Width Fiddling}\label{sec:bit_width}
As \ouralg works on the finite integer field, any model running on \ouralg should be quantized through quantization-aware training~\cite{nagel2022overcoming} or post-training quantization~\cite{liu2021post}. Considering GPU is designed for 64-bit floating point computing, it is necessary to discuss in what bit width a neural network should be quantized to, so that corresponding QNNs can be seamlessly embedded on the \ouralg framework. For the 64-bit floating point in IEEE 754 standard, the largest exact positive integer is $2^{53}$,
meaning that any computing results larger than it could induce error. Therefore, the 53-bit is the upper bound of all computing results in \ouralg, which is also confirmed in CryptGPU~\cite{tan2021cryptgpu}. With mature quantization strategies, many works demonstrated that networks with 16-bit precision~\cite{gupta2015deep} will not incur noticeable accuracy loss, compared with their 64/32-bit counterparts.
For 16-bit fixed point data, they can be rescaled to 16-bit integers, thus suited in the operating finite field of our \ouralg framework. 

Since all the computations on SSS finite field $\mathbb{F}_p$ must be reconstructable, e.g., $\mathsf{REC}(Conv(\llbracket x\rrbracket,\llbracket w\rrbracket))=Conv(x, w)$, we need to ensure the secret range stays in $\mathbb{F}_p$, indicating that the ring size $p$ must be carefully designed. For the multiplication, if both operands $x$ and $w$ are 16-bit, then the result is 32-bit; plus a preventive redundancy that allows the sum of $2^{13}$ multiplication results (which is large enough for convolution), the $p$ should be a prime number that is closest to $2^{45}$. Thus, we choose $p=35184372088777=2^{45}-55$ in the \ouralg framework for the 16-bit neural networks. Further, since each party will perform the convolution (and other operations) in the SSS format, i.e., $Conv(\llbracket x\rrbracket,\llbracket w\rrbracket)$, the overflow on the GPU computing should be considered. 
This is because calculations within the finite field $\mathbb{F}_p$ are not natively supported by GPU, lacking automatic modulo operations with respect to $p$. It is essential to ensure accurate representation of intermediate results prior to applying the modulo operation. For example, the multiplication on two shares $\textbf{A}, \textbf{B}\in\mathbb{F}_p$ will result in a 90-bit integer before modulo, which exceeds the representation range of the GPU 64-bit floating point numbers, but favors 128-bit computation.

{We follow the data decomposition strategy in \cite{tan2021cryptgpu} to propose corresponding strategy in \ouralg, performing the safe computation on the 64-bit floating-point GPU. Considering two numbers \textbf{A}, \textbf{B} $\in \mathbb{F}_p$, we first represent them as 46-bit numbers with zero-padding. Then, we split the data from the 23-th bit, i.e., $\textbf{A}=\textbf{A}_H * 2^{23}+\textbf{A}_L$ and $\textbf{B}=\textbf{B}_H * 2^{23}+\textbf{B}_L$, so that $\textbf{A}_H, \textbf{A}_L, \textbf{B}_H$, and $\textbf{B}_L$ are 23-bit.
The multiplication $\textbf{C}=(\textbf{A}\cdot\textbf{B})\ \text{mod}\ p$ is performed as
\begin{equation}
    \textbf{C}=\textbf{A}_H\textbf{B}_H2^{46} + \textbf{A}_H\textbf{B}_L2^{23} + \textbf{A}_L\textbf{B}_H2^{23}+\textbf{A}_L\textbf{B}_L
\end{equation}
Each term will be calculated separately and executed on modulo operation in parallel. 
We also provide a detailed computing procedure in Appendix D.
This computing procedure guarantees that all involved multiplications can produce integers in $\mathbb{F}_p$, without overflow on 64-bit floating-point GPU. Although one step of $\textbf{A}\cdot\textbf{B}$ is decomposed to multiple-step multiplications and additions, the advance of current GPU acceleration can well optimize this procedure without heavy overhead. Therefore, all the shares and data $\in\mathbb{F}_p$ in \ouralg framework are of the format of 23-bit splits.
However, we compose splits back to their 45-bit representations before communication and do decomposition again after data are received, to lighten the communication.}

{\textbf{Remark: A review on operation arrangement.} The SSS-Linear will output untruncated results, i.e., at most 45-bit that reaches the bit-width of $p$. If SSS-NonLinear is followed, then to make the multiplicative masking result ($x\times \beta$) recoverable in $\mathbb{F}_p$, there is no bit-width space for $\beta$ to do multiplication. Otherwise, we can further increase the bit-width of $p$, which might later require more splits to conduct the safe computation on GPU. Thus, truncation is preferred before the non-linear operation to reduce the bit-width of non-linear operation input to 16-bit. Plus the observations from Fig.~\ref{fig:communication} that \textbf{Linear-Truncation-NonLinear} order mostly has lower communication, we use this order as a default strategy for \ouralg.}

\subsection{Random Mask Generation}
The random masks in SSS-Truncation and SSS-NonLinear are critical and must be designed carefully to attain security and computation correctness simultaneously.

\textbf{Additive mask $\alpha$/$\widetilde{\alpha}$ generation.} The input $x$ of SSS-Truncation that is the output of SSS-Linear is the untruncated result of convolution/dense layer. As aforementioned, $x$'s bit-width should be 32 plus the logarithm of the number of additions, which is at most $2^{13}$ under our redundancy setting. Here $\alpha$ is just another addition to the summation result, so we still define the $\alpha$ bit-width as 32, regarding it as one of the $2^{13}$ redundancy. Par the requirement, $\alpha$ must be multiplies of the scaling factor $r$, $\alpha$ is generated following
\begin{equation*}
    \alpha\in\{\hat{\alpha}\in(0,2^{32}]\ |\ \hat{\alpha}\equiv 0 (\text{mod}\ r)\}
\end{equation*}
and $\widetilde{\alpha}=-\alpha/r=-e$. The cardinality of $\alpha$ and $\widetilde{\alpha}$ generation, i.e., the sampling space of $e$, is $\lfloor 2^{32}/r\rfloor$.
Especially, if the followed SSS-NonLinear operation includes AvgPooling, with kernel size $k_h\times k_w$, the condition of $\alpha$ is $\hat{\alpha}\equiv 0 (\text{mod}\ r\cdot k_h\cdot k_w)$. In this case, the cardinality of masking generation is reduced to $\lfloor 2^{32}/r/k_h/k_w\rfloor$.

\textbf{Multiplicative mask $\beta$/$\beta^{-1}$ generation.} Since SSS-NonLinear follows the truncation, the input $x$ now is 16-bit data. Therefore, the $\beta$ generation needs to satisfy the requirement that $x\times \beta<p/2$, i.e., making the multiplication result recoverable and the sign of $x$ not flipped. As we configure $p$ as a 45-bit large prime, the bit width of $\beta$ should be $28(=45-1-16)$, i.e., $\beta$ is generated following $\beta\in(0,2^{28}]$. Accordingly, $\beta^{-1}$ is determined as $\beta^{-1}\cdot \beta= 1$. The cardinality of $\beta$ generation is therefore $2^{28}$.

\section{\ouralg: Experimental Evaluation}
\label{sec:evaluation}

\subsection{Configuration}
\textbf{Environment.} We run our prototype on Amazon EC2 instances \texttt{g5.xlarge}, under the local area network (LAN) setting. Specifically, all the parties exist in the \texttt{us-east-1} (Northern Virginia) region, and each party (instance) is embedded with one NVIDIA A10G GPU with 24GB GPU memory. By GPU acceleration, we conduct all computations on GPU, keeping communication on CPU. All parties run in the Ubuntu 20.04 system with PyTorch 2.0.1 equipped. For the network configuration, we measured the latency of 0.1ms$\sim$0.2ms between parties, and the peak bandwidth of 1.25 GB/s. We set up this environment on the commercial cloud computing hardwares, to evaluate \ouralg in practical MLaaS use. While not exactly the same, we evaluate our \ouralg framework in the environment as close to previous GPU-based MPC works \cite{tan2021cryptgpu, kumar2020cryptflow} as possible.

\begin{table*}[t]
\caption{The secure inference analysis of \ouralg and SOTA MPC works, under the 3-party scenario. The execution time (in seconds) is averaged by 5 runs. The communication (Comm.) is measured as the total communication from all 3 parties. Plaintext is for the insecure inference of ANNs without any protection. All data are from \cite{tan2021cryptgpu} since CryptGPU slightly changed neural architectures and our tested architectures align with CryptGPU for fair comparison. Values in parentheses are estimated.}
\resizebox{\textwidth}{!}{
\begin{tabular}{@{}l|cccccccccc@{}}
\toprule[.5mm]
 & \multicolumn{2}{c}{LeNet (MNIST)} & \multicolumn{2}{c}{AlexNet (Cifar-10)} & \multicolumn{2}{c}{VGG-16 (Cifar-10)} & \multicolumn{2}{c}{AlexNet (TI)} & \multicolumn{2}{c}{VGG-16 (TI)} \\ 
 & Time (s) & Comm. (MB) & Time (s) & Comm. (MB) & Time (s) & Comm. (MB) & Time (s) & Comm. (MB) & Time (s) & Comm. (MB) \\\midrule
Falcon \cite{wagh2020falcon} & 0.038 & 2.29 & 0.11 & 4.02 & 1.44 & 40.45 & 0.34 & 16.23 & 8.61 & 161.71 \\
CryptGPU \cite{tan2021cryptgpu}& 0.38 & 3.00 & 0.91 & 2.43 & 2.14 & 56.2 & 0.95 & 13.97 & 2.30 & 224.5 \\
\textbf{\ouralg} & \textbf{0.027} & \textbf{1.34} & \textbf{0.113} & \textbf{1.19} & \textbf{0.363} & \textbf{26.04} & \textbf{0.150} & \textbf{6.31} & \textbf{0.486} & \textbf{104.11} \\
Plaintext & 0.0007 & $-$ & 0.0012 & $-$ & 0.0024 & $-$ & 0.0012 & $-$ & 0.0024 & $-$ \\\hline\hline
 & \multicolumn{2}{c}{AlexNet (ImageNet)} & \multicolumn{2}{c}{VGG-16 (ImageNet)} & \multicolumn{2}{c}{ResNet-50 (ImageNet)} & \multicolumn{2}{c}{ResNet-101 (ImageNet)} & \multicolumn{2}{c}{ResNet-152 (ImageNet)} \\
 & Time (s) & Comm. (GB) & Time (s) & Comm. (GB) & Time (s) & Comm. (GB) & Time (s) & Comm. (GB) & Time (s) & Comm. (GB) \\\midrule
CrypTFlow \cite{kumar2020cryptflow}& $-$ & $-$ & $-$ & $-$ & 25.9 & 6.9 & (40) & (10.5) & (60) & (14.5) \\
CryptGPU \cite{tan2021cryptgpu}& 1.52 & 0.24 & 9.44 & 2.75 & 9.31 & 3.08 & 17.62 & 4.64 & 25.77 & 6.56 \\
\textbf{\ouralg} & \textbf{0.217} & \textbf{0.057} & \textbf{2.78} & \textbf{1.27} & \textbf{1.96} & \textbf{1.05} & \textbf{3.19} & \textbf{1.56} & \textbf{4.52} & \textbf{2.18} \\
Plaintext & 0.0013 & $-$ & 0.0024 & $-$ & 0.011 & $-$ & 0.021 & $-$ & 0.031 & $-$ \\ \bottomrule[.5mm]
\end{tabular}}
\label{tab:performance_comparison}
\end{table*}

\textbf{Datasets.} We evaluate \ouralg on modern classification tasks: MNIST \cite{mnist}, Cifar-10 \cite{cifar10}, Tiny ImageNet~\cite{tinyimagenet}, and ImageNet~\cite{imagenet2009}. From small to large scale, MNIST has $1\times28\times28$-sized inputs and 10 classes, Cifar-10 has $3\times32\times32$-sized inputs and 10 classes, Tiny ImageNet contains $3\times64\times64$ RGB inputs with 200 classes, and ImageNet as the largest classification task, has $3\times224\times224$ RGB inputs with 1000 classes. While the small-scale tasks have been widely investigated, large and difficult tasks for privacy-preserving ML such as ImageNet have only been evaluated in \cite{tan2021cryptgpu, kumar2020cryptflow}. Since we consider 16-bit input, all these tasks' inputs are extended to 16 bits with zero padding.

\textbf{Artifical Neural Networks.} We evaluate \ouralg on models scaled from shallow ones to large, which include LeNet~\cite{lenet}, AlexNet~\cite{alexnet}, VGG~\cite{vgg}, and ResNet~\cite{resnet}. The activation functions of all networks are altered to ReLU to fit in our framework. Also, we conduct other tiny architecture adjustments as mentioned in~\cite{tan2021cryptgpu}, for fair comparison.

\subsection{Inference Analysis}
We share a comprehensive performance analysis on \ouralg and other SOTA MPC works, e.g., CryptGPU \cite{tan2021cryptgpu}, CrypTFlow \cite{kumar2020cryptflow}, Falcon \cite{wagh2020falcon}. \ouralg is proposed for the GPU acceleration (under PyTorch) with our data decomposition technique. Although \ouralg can also run on CPU platforms as Falcon, the decomposition can be removed if deploying \ouralg under numerical computing libraries, such as SciPy that supports direct 128-bit computation, further reducing the computation complexity.

\subsubsection{Inference Comparison}
We demonstrate the overall performance analysis in Tab.~\ref{tab:performance_comparison}, inferring one image (batch= 1). Note that Falcon is executed on CPU platforms, while other frameworks (including Plaintext) are GPU-accelerated. For communication, our \ouralg framework achieves the lowest volume on all benchmarks. \ouralg achieves the highest $76\%$ less communication on AlexNet (ImageNet), and even in the lowest case on AlexNet (Cifar-10), \ouralg still has $51\%$ communication reduction, compared with CryptGPU. The significant improvement in communication efficiency mainly lies in the lightweight design of our SSS-NonLinear operation, as discussed in Tab.~\ref{tab:commcomplexity}. On the other hand, CryptGPU and Falcon employ analogous ABY$^3$ protocols, resulting in similar communication volumes. Regarding execution time, \ouralg attains a speed-up ranging from $\times 3.4$ to $\times 14$ over CryptGPU. This acceleration is notably pronounced in smaller to moderate tasks, such as MNIST and Cifar-10, where it varies between $\times 5.9$ and $\times 14$. However, this increase is less marked in larger tasks and models, e.g., ResNet on ImageNet, which approximates $\times 5$. Still, even with significant improvement from SOTA frameworks, \ouralg is still hundreds of times slower than the execution time of unprotected DNNs (plaintext), appealing to the necessity of more efficient paradigms on the secret generation/reconstruction and secure computation.

\begin{table}[t]
\caption{Accuracy comparison between the full-precision and the 16-bit quantized version conducted by \ouralg framework. M--MNIST, TI--Tiny ImageNet, I--ImageNet.}
\resizebox{\linewidth}{!}{
\begin{tabular}{c|ccccc}
\toprule[.5mm]
 & LeNet(MNIST) & AlexNet(TI) & VGG(TI) & ResNet50(I) & ResNet101(I) \\\midrule
Full & $99.4\%$ & $46.3\%$ & $52.1\%$ & $77.7\%$ & $79.2\%$ \\
16-bit & $99.3\%$ & $46.2\%$ & $52.1\%$ & $77.7\%$ & $79.2\%$ \\\bottomrule[.5mm]
\end{tabular}}
\label{tab:accuracy_analysis}
\end{table}

While MPC frameworks could modify the backbone neural architecture, previous works \cite{tan2021cryptgpu, kumar2020cryptflow} have proved that slight modifications, e.g., on quantization and pooling, will not decisively steer accuracy in the worse direction, but lead to comparable accuracy against the orginal DNN architectures. For completeness, we also demonstrate in Tab.~\ref{tab:accuracy_analysis} the comparison before and after the modifications, i.e., 16-bit quantization, on 5 selected tasks from small scale (LeNet on MNIST) to large scale (ResNet-101 on ImageNet). 
There is no obvious accuracy degradation when quantizing the DNN from full-precision floating-point to 16-bit fixed-point.

\begin{table}[t]
\caption{Batched inference of \ouralg on Cifar-10 with size $bsz$. Time is in seconds and communication is in GB. }
\label{tab:batch_cifar}
\centering
\resizebox{.85\linewidth}{!}{
\begin{tabular}{@{}cc|cccc@{}}
\toprule[.5mm]
 &  & \multicolumn{2}{c}{$bsz=1$} & \multicolumn{2}{c}{$bsz=64$} \\
 &  & Time & Comm. & Time & Comm. \\\midrule
\multirow{2}{*}{CryptGPU} & AlexNet & 0.91 & 0.002 & 1.09 & 0.16 \\
 & VGG-16 & 2.14 & 0.056 & 11.76 & 3.60 \\\hline
\multirow{2}{*}{\textbf{\ouralg}} & AlexNet & \textbf{0.113} & \textbf{0.001} & \textbf{0.262} & \textbf{0.076} \\
 & VGG-16 & \textbf{0.363} & \textbf{0.026} & \textbf{3.73} & \textbf{1.67} \\\bottomrule[.5mm]
\end{tabular}}
\end{table}

\subsubsection{Batch Comparison}
We evaluate the batched inference on small-scale (Cifar-10 in Tab.~\ref{tab:batch_cifar}) and large-scale (ImageNet in Tab.~\ref{tab:batch_imagenet}) tasks, by comparing with CryptGPU that also applied batch inference on GPU acceleration. For communication, the volume is always linear to the batch size, because the shares of all model parameters are distinctively prepared for each image input to ensure data privacy. This is a major difference between MPC models and plaintext models which have identical parameters (i.e., weights) for all inputs. This observation also aligns with CryptGPU. For the execution time, while \ouralg can always achieve better performance than CryptGPU, there are interesting observations can be obtained. For the small-scale task (e.g., Cifar-10), we find similarity between \ouralg and CryptGPU, that batch inference can greatly amortize the inference time; from 1 image to 64 images, the execution time only increases $\times 2.32$(AlexNet) and $\times 10.3$(VGG-16) times. However, while this amortization still exists in CryptGPU for the large task, ImageNet, there is no obvious advantage of batch inference for \ouralg. Specifically, the inference time of all ResNet models under \ouralg framework has $\times 8$ increment, by increasing the number of images from 1 to 8. This could come from the heavy computation and communication on degree reduction in SSS-Linear, which requires share generation and reconstruction. 

\begin{table}[t]
\caption{Batched inference of \ouralg on ImageNet with size $bsz$. Time is in seconds and communication is in GB. }
\label{tab:batch_imagenet}
\centering
\resizebox{.85\linewidth}{!}{
\begin{tabular}{@{}cc|cccc@{}}
\toprule[.5mm]
 &  & \multicolumn{2}{c}{$bsz=1$} & \multicolumn{2}{c}{$bsz=8$} \\
 &  & Time & Comm. & Time & Comm. \\\midrule
\multirow{3}{*}{CryptGPU} & ResNet-50 & 9.31 & 3.08 & 42.99 & 24.7 \\
 & ResNet-101 & 17.62 & 4.64 & 72.99 & 37.2 \\
  & ResNet-152 & 25.77 & 6.56 & 105.20 & 52.5\\\hline
\multirow{3}{*}{\textbf{\ouralg}} & ResNet-50 & \textbf{1.96} & \textbf{1.05} & \textbf{16.92} & \textbf{8.39} \\
 & ResNet-101 & \textbf{3.19} & \textbf{1.56} & \textbf{24.22} & \textbf{12.5} \\
  & ResNet-152 & \textbf{4.52} & \textbf{2.18} & \textbf{33.60} & \textbf{17.5} \\\bottomrule[.5mm]
\end{tabular}}
\end{table}

\begin{figure}[t]
    \centering
    \includegraphics[width=.8\linewidth]{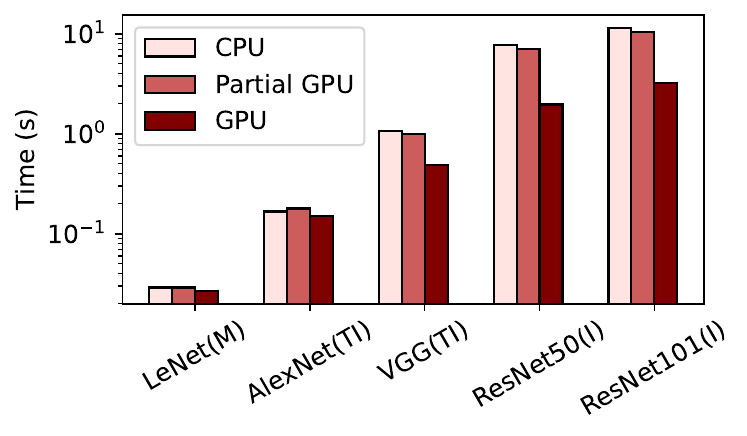}
    \caption{The execution time of \ouralg frameworks on selected benchmarks, regarding pure CPU computing, partial-GPU accelerating, and full-GPU accelerating. Time is in log scale. M--MNIST, TI--Tiny ImageNet, I--ImageNet.}
    \label{fig:gpu_acceleration}
\end{figure}

\subsubsection{GPU Acceleration}\label{sec:gpu_acceleration}
As computation is another major concern in MPC-based frameworks, the powerful floating-point processing ability in GPU significantly accelerates the computation procedure. On the other hand, since current commercial GPUs (at least for ours and backends in CryptGPU~\cite{tan2021cryptgpu})
are not able to directly perform the communication, which means the communication must be conducted on the CPU side. Frequent data transmission between GPU and peripheral memories (connected to Ethernet) involves extra time overhead, despite the GPU's acceleration. Therefore, a trade-off could exist on whether the computation between two consecutive communications is accelerated by GPU. 

To employ this investigation, we configure three scenarios: pure CPU computing, only $Conv2D$/$Dense$ accelerated by GPU (Step \colorcircled{black}{white}{1} of SSS-Linear in Fig.~\ref{fig:SSSOpeartion}), and all computations accelerated by GPU. We explore this trade-off on tasks of small scale (LeNet on MNIST), moderate scale (AlexNet on TI, VGG-16 on TI), and large scale (ResNet-50 on ImageNet, ResNet-101 on ImageNet), respectively. As the communication stays the same regardless of the computation scheduling, we only show the inference time in Fig.~\ref{fig:gpu_acceleration}. For small to moderate benchmarks, the partial GPU acceleration could be even worse than pure CPU computing, e.g., AlexNet on TI; this is mainly because our selected CPU is powerful enough to handle lightweight model execution, so the benefit from GPU acceleration is inferior to the overhead from data exchange between CPU and GPU. However, the GPU acceleration is significant for large benchmarks, e.g., ImageNet, even only accelerating the SSS-Linear operation. Besides, the acceleration on other operations is also obvious (as in the full-GPU accelerating case), making the execution time much shorter than the case when the CPU participates in partial computation, even considering the data exchange overhead.
Par this observation, we utilize GPU to accelerate all the computations in \ouralg framework.

\subsubsection{Varying Party Numbers}
While SOTA MPC works are dedicated to a specific number of parties, e.g., 2PC \cite{riazi2019xonn} and 3PC \cite{mohassel2018aby3}, \ouralg can be easily set up as an environment with arbitrary parties, as long as satisfying the scheme requirement, e.g., $n\geq 2k-1$ for $(k,n)$-SSS. The secure inference performance of 3PC with $(2,3)$-SSS scheme and 5PC with $(3,5)$-SSS scheme is depicted in Tab.~\ref{tab:vary_party}. By scaling up to 5 parties, one active party and one passive party are added. The communication volume increases by $\times 2.5$ on all benchmarks. This corresponds to our complexity analysis in Tab.~\ref{tab:commcomplexity}, that the communication on SSS-Linear operation increases quadratically against the party number. One sequence of Linear-Truncation-NonLinear operation involves communication $(3k^2+4k-4)$, thus the theoretical communication ratio of 5PC($k=3$) over 3PC($k=2$) is $2.2$, which is close to our observation here. For the execution time, the increment from 3PC to 5PC is only $26\%-55\%$ on small tasks (MNIST and TI) and $74\%-86\%$ on large tasks (ImageNet). Although all the model-related computation is independent of the party number, the SSS-related computations (e.g., share generation and secret reconstruction) are highly dependent on the involved compute parties. Besides, the waiting time for communication will also increase regarding more parties.

\begin{table}[t]
\caption{Inference performance of \ouralg framework by varying the number of parties, i.e., 3PC and 5PC. Time is in seconds and communication is in GB.}
\label{tab:vary_party}
\centering
\resizebox{.9\linewidth}{!}{
\begin{tabular}{@{}c|cccc@{}}
\toprule[.5mm]
& \multicolumn{2}{c}{$\#=3$} & \multicolumn{2}{c}{$\#=5$} \\
& Time & Comm. & Time & Comm. \\\midrule
LeNet(MNIST) & 0.027 & 0.001 & 0.042 & 0.003 \\
AlexNet(TI) & 0.150 & 0.006 & 0.189 & 0.016 \\
VGG-16(TI) & 0.486 & 0.104 & 0.664 & 0.259 \\
ResNet-50(ImageNet) & 1.96 & 1.05 & 3.64 & 2.61 \\
ResNet-101(ImageNet) & 3.19 & 1.56 & 5.54 & 3.85 \\\bottomrule[.5mm]
\end{tabular}}
\end{table}

\begin{figure}[t]
    \centering
\subfigure{\centering
    \includegraphics[width=\linewidth]{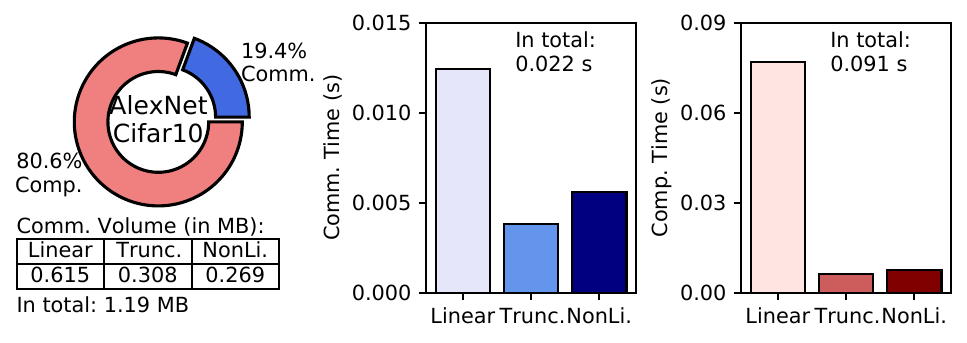}}
\subfigure{\centering
    \includegraphics[width=\linewidth]{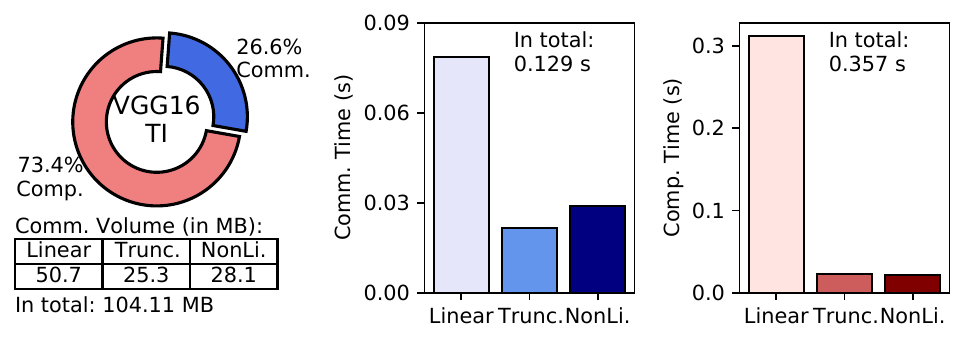}}
\subfigure{\centering
    \includegraphics[width=\linewidth]{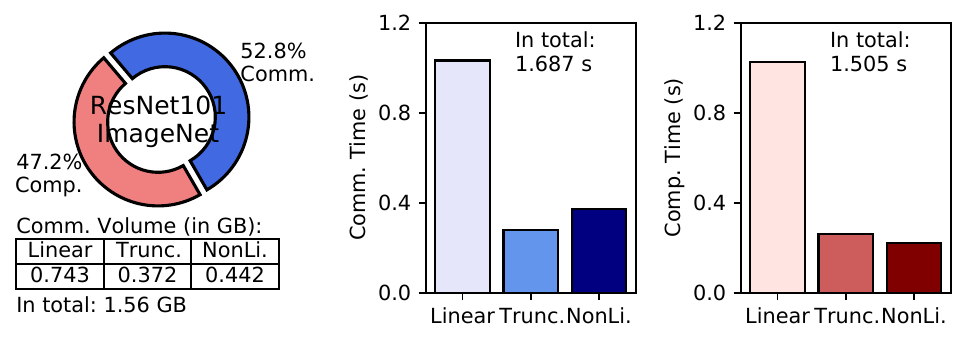}}
\caption{The performance decomposition of \ouralg framework on small-scale (AlexNet on Cifar-10), moderate-scale (VGG-16 on Tiny ImageNet), and large-scale (ResNet-101 on ImageNet) DNN inference. We demonstrate the computation time (Comp.) and communication time (Comm.) in pie charts and bar plots, and the communication volume in tables, of three SSS operations. Trunc. -- Truncation, NonLi. -- NonLinear.}
    \label{fig:ablation_study}
\end{figure}

\subsection{Ablation Study}\label{sec:ablation}
We explore the time and communication overhead of distinctive SSS operations inside DNNs secured by \ouralg. This will help to investigate the consumption during model inference. Selecting one small-scale task (AlexNet on Cifar10), one moderate-scale task (VGG-16 on Tiny ImageNet), and one large-scale task (ResNet-101 on ImageNet), we provide the performance decomposition on all SSS operations, in Fig.~\ref{fig:ablation_study}. The performance decomposition of other tasks is provided in Appendix E. Along the task scale, one significant observation is that the communication time on larger tasks (e.g., from Cifar10 to ImageNet) takes more ratio of the entire execution time, from $19.4\%$ to $52.8\%$. This arises from the fact that although computation can be parallelized within the confines of GPU resources, inter-party communication is constrained by LAN settings, where data must be sequentially transferred through P2P channels. As ablation, SSS-linear consumes most time on both communication and computation, since the \textit{Conv2D/Dense} layers are computing-intensive, highlighting the necessary of GPU acceleration, as discussed in Sec.~\ref{sec:gpu_acceleration}. In terms of communication volume, SSS-Linear requires approximately double the communication compared to SSS-Truncation and SSS-NonLinear. This observation aligns with the communication time analysis depicted in the bar plots. This trend remains consistent across all tasks, attributable to the fact that nearly all layers in any given model follow the operational sequence of Linear-Truncation-NonLinear. Additionally, this analysis corresponds with the communication complexity in Tab.~\ref{tab:commcomplexity}, where NonLinear operations consume substantially less communication time compared to Linear operations. This contrasts with previous MPC frameworks, such as \cite{tan2021cryptgpu}, where NonLinear operations (e.g., ReLU) incur communication time comparable to Linear operations. 

\subsection{\ouralg in Practical Network Setup}
To put the secure model inference under \ouralg framework forward to practical use, we adopt the wide area network (WAN) setting to investigate the execution time. Specifically, we initialize three parties for the $(2,3)$-SSS scheme, where one is in Northern Virginia (A10G GPU) as elite, one is in Ohio (A10G GPU) as active, and the other is in Eastern Massachusetts (RTX3070 GPU) as passive. The measured latency between them is around (9.5ms, 12.3ms, 21.6ms). We demonstrate the execution time of selected benchmarks in Tab. \ref{tab:wan}. For small-scale tasks, such as MNIST and Tiny ImageNet, the execution time increases around $\times 10$ under the realistic WAN setting; yet for large-scale ImageNet tasks, the execution consumes more than $\times 20$ time. Although our \ouralg framework could outperform other frameworks under the WAN setting, e.g., LeNet(MNIST) consumes 3s in Falcon (with 70ms latency) yet only 0.35s in \ouralg (with 21ms latency), there are still gaps between the MPC environmental evaluation and practical use. For example in the ResNet-101 test, the plaintext unprotected inference time is 0.021s, while the experimental LAN setting consumes $\times 152$ more overhead (3.19s), and the practical WAN setting even takes $\times 3261$ more overhead (68.49s). This is also a critical motivation for further revolutionary computing/communication paradigms in MPC development.

\begin{table}[t]
\caption{The secure inference time for 3PC \ouralg framework under WAN setting. M---MNIST, TI---Tiny ImageNet, and I---ImageNet. }
    \label{tab:wan}
    \centering
    \resizebox{\linewidth}{!}{
    \begin{tabular}{c|ccccc}
    \toprule[.5mm]
         & LeNet(M) & AlexNet(TI) & VGG(TI)& ResNet50(I)& ResNet101(I) \\\midrule
       Time (s)  & 0.351 &0.982 &5.40 &43.64 & 68.49\\\bottomrule[.5mm]
    \end{tabular}}
\end{table}

\section{Conclusion}
In this work, we propose the first Shamir's secret sharing) based framework, \ouralg, to conduct the multi-party computation on privacy-preserving machine learing service. Unlike previous additive secret sharing based approaches, \ouralg provides superior scalability and robustness for modern MPC application, under the SSS scheme. Moreover, we introduce novel secure comparison strategies, which substantially reduce the communication complexity on non-linear operations during DNN inference. Other secure primitives and operations are introduced as well to complete our \ouralg framework. The evaluation on \ouralg demonstrates the speed-up over previous MPC frameworks, and for the first time, we evaluate \ouralg as a MPC framework on five-party scenario in real-world commercial cloud and networking platforms. 


\bibliographystyle{IEEEtran}
\bibliography{main.bbl}

\begin{thebibliography}{10}
\providecommand{\url}[1]{#1}
\csname url@samestyle\endcsname
\providecommand{\newblock}{\relax}
\providecommand{\bibinfo}[2]{#2}
\providecommand{\BIBentrySTDinterwordspacing}{\spaceskip=0pt\relax}
\providecommand{\BIBentryALTinterwordstretchfactor}{4}
\providecommand{\BIBentryALTinterwordspacing}{\spaceskip=\fontdimen2\font plus
\BIBentryALTinterwordstretchfactor\fontdimen3\font minus
  \fontdimen4\font\relax}
\providecommand{\BIBforeignlanguage}[2]{{%
\expandafter\ifx\csname l@#1\endcsname\relax
\typeout{** WARNING: IEEEtran.bst: No hyphenation pattern has been}%
\typeout{** loaded for the language `#1'. Using the pattern for}%
\typeout{** the default language instead.}%
\else
\language=\csname l@#1\endcsname
\fi
#2}}
\providecommand{\BIBdecl}{\relax}
\BIBdecl

\bibitem{weng2022mlaas}
Q.~Weng, W.~Xiao, Y.~Yu, W.~Wang, C.~Wang, J.~He, Y.~Li, L.~Zhang, W.~Lin, and
  Y.~Ding, ``Mlaas in the wild: Workload analysis and scheduling in large-scale
  heterogeneous gpu clusters,'' in \emph{19th USENIX Symposium on Networked
  Systems Design and Implementation (NSDI 22)}, 2022, pp. 945--960.

\bibitem{li2021survey}
Q.~Li, Z.~Wen, Z.~Wu, S.~Hu, N.~Wang, Y.~Li, X.~Liu, and B.~He, ``A survey on
  federated learning systems: Vision, hype and reality for data privacy and
  protection,'' \emph{IEEE Transactions on Knowledge and Data Engineering},
  2021.

\bibitem{joshi2022comparative}
B.~Joshi, B.~Joshi, A.~Mishra, V.~Arya, A.~K. Gupta, and D.~Perakovi{\'c}, ``A
  comparative study of privacy-preserving homomorphic encryption techniques in
  cloud computing,'' \emph{International Journal of Cloud Applications and
  Computing (IJCAC)}, vol.~12, no.~1, pp. 1--11, 2022.

\bibitem{zhao2019secure}
C.~Zhao, S.~Zhao, M.~Zhao, Z.~Chen, C.-Z. Gao, H.~Li, and Y.-a. Tan, ``Secure
  multi-party computation: theory, practice and applications,''
  \emph{Information Sciences}, vol. 476, pp. 357--372, 2019.

\bibitem{tan2021cryptgpu}
S.~Tan, B.~Knott, Y.~Tian, and D.~J. Wu, ``Cryptgpu: Fast privacy-preserving
  machine learning on the gpu,'' in \emph{2021 IEEE Symposium on Security and
  Privacy (SP)}.\hskip 1em plus 0.5em minus 0.4em\relax IEEE, 2021, pp.
  1021--1038.

\bibitem{knott2021crypten}
B.~Knott, S.~Venkataraman, A.~Hannun, S.~Sengupta, M.~Ibrahim, and L.~van~der
  Maaten, ``Crypten: Secure multi-party computation meets machine learning,''
  \emph{Advances in Neural Information Processing Systems}, vol.~34, pp.
  4961--4973, 2021.

\bibitem{mishra2020delphi}
P.~Mishra, R.~Lehmkuhl, A.~Srinivasan, W.~Zheng, and R.~A. Popa, ``Delphi: A
  cryptographic inference system for neural networks,'' in \emph{Proceedings of
  the 2020 Workshop on Privacy-Preserving Machine Learning in Practice}, 2020,
  pp. 27--30.

\bibitem{wagh2020falcon}
S.~Wagh, S.~Tople, F.~Benhamouda, E.~Kushilevitz, P.~Mittal, and T.~Rabin,
  ``Falcon: Honest-majority maliciously secure framework for private deep
  learning,'' \emph{arXiv preprint arXiv:2004.02229}, 2020.

\bibitem{kumar2020cryptflow}
N.~Kumar, M.~Rathee, N.~Chandran, D.~Gupta, A.~Rastogi, and R.~Sharma,
  ``Cryptflow: Secure tensorflow inference,'' in \emph{2020 IEEE Symposium on
  Security and Privacy (SP)}.\hskip 1em plus 0.5em minus 0.4em\relax IEEE,
  2020, pp. 336--353.

\bibitem{beaver1992efficient}
D.~Beaver, ``Efficient multiparty protocols using circuit randomization,'' in
  \emph{Advances in Cryptology—CRYPTO’91: Proceedings 11}.\hskip 1em plus
  0.5em minus 0.4em\relax Springer, 1992, pp. 420--432.

\bibitem{araki2016high}
T.~Araki, J.~Furukawa, Y.~Lindell, A.~Nof, and K.~Ohara, ``High-throughput
  semi-honest secure three-party computation with an honest majority,'' in
  \emph{Proceedings of the 2016 ACM SIGSAC Conference on Computer and
  Communications Security}, 2016, pp. 805--817.

\bibitem{riazi2019xonn}
M.~S. Riazi, M.~Samragh, H.~Chen, K.~Laine, K.~Lauter, and F.~Koushanfar,
  ``Xonn:xnor-based oblivious deep neural network inference,'' in \emph{28th
  USENIX Security Symposium (USENIX Security 19)}, 2019, pp. 1501--1518.

\bibitem{mohassel2017secureml}
P.~Mohassel and Y.~Zhang, ``Secureml: A system for scalable privacy-preserving
  machine learning,'' in \emph{2017 IEEE symposium on security and privacy
  (SP)}.\hskip 1em plus 0.5em minus 0.4em\relax IEEE, 2017, pp. 19--38.

\bibitem{mohassel2018aby3}
P.~Mohassel and P.~Rindal, ``Aby3: A mixed protocol framework for machine
  learning,'' in \emph{Proceedings of the 2018 ACM SIGSAC conference on
  computer and communications security}, 2018, pp. 35--52.

\bibitem{chaudhari2019trident}
H.~Chaudhari, R.~Rachuri, and A.~Suresh, ``Trident: Efficient 4pc framework for
  privacy preserving machine learning,'' \emph{arXiv preprint
  arXiv:1912.02631}, 2019.

\bibitem{byali2019flash}
M.~Byali, H.~Chaudhari, A.~Patra, and A.~Suresh, ``Flash: Fast and robust
  framework for privacy-preserving machine learning,'' \emph{Cryptology ePrint
  Archive}, 2019.

\bibitem{yao1986generate}
A.~C.-C. Yao, ``How to generate and exchange secrets,'' in \emph{27th annual
  symposium on foundations of computer science (Sfcs 1986)}.\hskip 1em plus
  0.5em minus 0.4em\relax IEEE, 1986, pp. 162--167.

\bibitem{liu2017oblivious}
J.~Liu, M.~Juuti, Y.~Lu, and N.~Asokan, ``Oblivious neural network predictions
  via minionn transformations,'' in \emph{Proceedings of the 2017 ACM SIGSAC
  conference on computer and communications security}, 2017, pp. 619--631.

\bibitem{furukawa2019two}
J.~Furukawa and Y.~Lindell, ``Two-thirds honest-majority mpc for malicious
  adversaries at almost the cost of semi-honest,'' in \emph{Proceedings of the
  2019 ACM SIGSAC Conference on Computer and Communications Security}, 2019,
  pp. 1557--1571.

\bibitem{stadler1996publicly}
M.~Stadler, ``Publicly verifiable secret sharing,'' in \emph{International
  Conference on the Theory and Applications of Cryptographic Techniques}.\hskip
  1em plus 0.5em minus 0.4em\relax Springer, 1996, pp. 190--199.

\bibitem{shamir1979share}
A.~Shamir, ``How to share a secret,'' \emph{Communications of the ACM},
  vol.~22, no.~11, pp. 612--613, 1979.

\bibitem{ball2016garbling}
M.~Ball, T.~Malkin, and M.~Rosulek, ``Garbling gadgets for boolean and
  arithmetic circuits,'' in \emph{Proceedings of the 2016 ACM SIGSAC Conference
  on Computer and Communications Security}, 2016, pp. 565--577.

\bibitem{ball2019garbled}
M.~Ball, B.~Carmer, T.~Malkin, M.~Rosulek, and N.~Schimanski, ``Garbled neural
  networks are practical,'' \emph{Cryptology ePrint Archive}, 2019.

\bibitem{diffie2022new}
W.~Diffie and M.~E. Hellman, ``New directions in cryptography,'' in
  \emph{Democratizing Cryptography: The Work of Whitfield Diffie and Martin
  Hellman}, 2022, pp. 365--390.

\bibitem{kamara2011outsourcing}
S.~Kamara, P.~Mohassel, and M.~Raykova, ``Outsourcing multi-party
  computation,'' \emph{Cryptology ePrint Archive}, 2011.

\bibitem{liu2019survey}
Y.~Liu, B.~Zhao, P.~Zhao, P.~Fan, and H.~Liu, ``A survey: Typical security
  issues of software-defined networking,'' \emph{China Communications},
  vol.~16, no.~7, pp. 13--31, 2019.

\bibitem{pearce2013virtualization}
M.~Pearce, S.~Zeadally, and R.~Hunt, ``Virtualization: Issues, security
  threats, and solutions,'' \emph{ACM Computing Surveys (CSUR)}, vol.~45,
  no.~2, pp. 1--39, 2013.

\bibitem{xiao2016one}
Y.~Xiao, X.~Zhang, Y.~Zhang, and R.~Teodorescu, ``One bit flips, one cloud
  flops:cross-vm row hammer attacks and privilege escalation,'' in \emph{25th
  USENIX security symposium (USENIX Security 16)}, 2016, pp. 19--35.

\bibitem{ben2019completeness}
M.~Ben-Or, S.~Goldwasser, and A.~Wigderson, ``Completeness theorems for
  non-cryptographic fault-tolerant distributed computation,'' in
  \emph{Providing Sound Foundations for Cryptography: On the Work of Shafi
  Goldwasser and Silvio Micali}, 2019, pp. 351--371.

\bibitem{grunwald1942theory}
G.~Gr{\"u}nwald, ``On the theory of interpolation,'' 1942.

\bibitem{lidl1994introduction}
R.~Lidl and H.~Niederreiter, \emph{Introduction to finite fields and their
  applications}.\hskip 1em plus 0.5em minus 0.4em\relax Cambridge university
  press, 1994.

\bibitem{watanabe2015secrecy}
T.~Watanabe, K.~Iwamura, and K.~Kaneda, ``Secrecy multiplication based on a (k,
  n)-threshold secret-sharing scheme using only k servers,'' in \emph{Computer
  Science and its Applications: Ubiquitous Information Technologies}.\hskip 1em
  plus 0.5em minus 0.4em\relax Springer, 2015, pp. 107--112.

\bibitem{shingu2016secrecy}
T.~Shingu, K.~Iwamura, and K.~Kaneda, ``Secrecy computation without changing
  polynomial degree in shamir's (k, n) secret sharing scheme.'' in
  \emph{DCNET}, 2016, pp. 89--94.

\bibitem{abraham2020blinder}
I.~Abraham, B.~Pinkas, and A.~Yanai, ``Blinder--scalable, robust anonymous
  committed broadcast,'' in \emph{Proceedings of the 2020 ACM SIGSAC Conference
  on Computer and Communications Security}, 2020, pp. 1233--1252.

\bibitem{yandamuri2021communication}
S.~Yandamuri, I.~Abraham, K.~Nayak, and M.~K. Reiter, ``Communication-efficient
  bft protocols using small trusted hardware to tolerate minority corruption,''
  \emph{Cryptology ePrint Archive}, 2021.

\bibitem{patra2020blaze}
A.~Patra and A.~Suresh, ``Blaze: blazing fast privacy-preserving machine
  learning,'' \emph{arXiv preprint arXiv:2005.09042}, 2020.

\bibitem{nagel2022overcoming}
M.~Nagel, M.~Fournarakis, Y.~Bondarenko, and T.~Blankevoort, ``Overcoming
  oscillations in quantization-aware training,'' \emph{arXiv preprint
  arXiv:2203.11086}, 2022.

\bibitem{liu2021post}
Z.~Liu, Y.~Wang, K.~Han, W.~Zhang, S.~Ma, and W.~Gao, ``Post-training
  quantization for vision transformer,'' \emph{Advances in Neural Information
  Processing Systems}, vol.~34, pp. 28\,092--28\,103, 2021.

\bibitem{gupta2015deep}
S.~Gupta, A.~Agrawal, K.~Gopalakrishnan, and P.~Narayanan, ``Deep learning with
  limited numerical precision,'' in \emph{International conference on machine
  learning}.\hskip 1em plus 0.5em minus 0.4em\relax PMLR, 2015, pp. 1737--1746.

\bibitem{mnist}
Y.~{Lecun}, L.~{Bottou}, Y.~{Bengio}, and P.~{Haffner}, ``Gradient-based
  learning applied to document recognition,'' \emph{Proceedings of the IEEE},
  vol.~86, no.~11, pp. 2278--2324, 1998.

\bibitem{cifar10}
A.~Krizhevsky, G.~Hinton \emph{et~al.}, ``Learning multiple layers of features
  from tiny images,'' 2009.

\bibitem{tinyimagenet}
Y.~Le and X.~Yang, ``Tiny imagenet visual recognition challenge,'' \emph{CS
  231N}, vol.~7, no.~7, p.~3, 2015.

\bibitem{imagenet2009}
J.~Deng, W.~Dong, R.~Socher, L.-J. Li, K.~Li, and L.~Fei-Fei, ``Imagenet: A
  large-scale hierarchical image database,'' in \emph{2009 IEEE conference on
  computer vision and pattern recognition}.\hskip 1em plus 0.5em minus
  0.4em\relax Ieee, 2009, pp. 248--255.

\bibitem{lenet}
Y.~LeCun, L.~Bottou, Y.~Bengio, and P.~Haffner, ``Gradient-based learning
  applied to document recognition,'' \emph{Proceedings of the IEEE}, vol.~86,
  no.~11, pp. 2278--2324, 1998.

\bibitem{alexnet}
A.~Krizhevsky, I.~Sutskever, and G.~E. Hinton, ``Imagenet classification with
  deep convolutional neural networks,'' \emph{Advances in neural information
  processing systems}, vol.~25, 2012.

\bibitem{vgg}
K.~Simonyan and A.~Zisserman, ``Very deep convolutional networks for
  large-scale image recognition,'' \emph{arXiv preprint arXiv:1409.1556}, 2014.

\bibitem{resnet}
K.~He, X.~Zhang, S.~Ren, and J.~Sun, ``Deep residual learning for image
  recognition,'' in \emph{Proceedings of the IEEE conference on computer vision
  and pattern recognition}, 2016, pp. 770--778.

\end{thebibliography}

\section*{Appendix}

\subsection*{A. Intuitive Example for Convolution Output}
\begin{figure}[t]
    \centering
    \includegraphics[width=.7\linewidth]{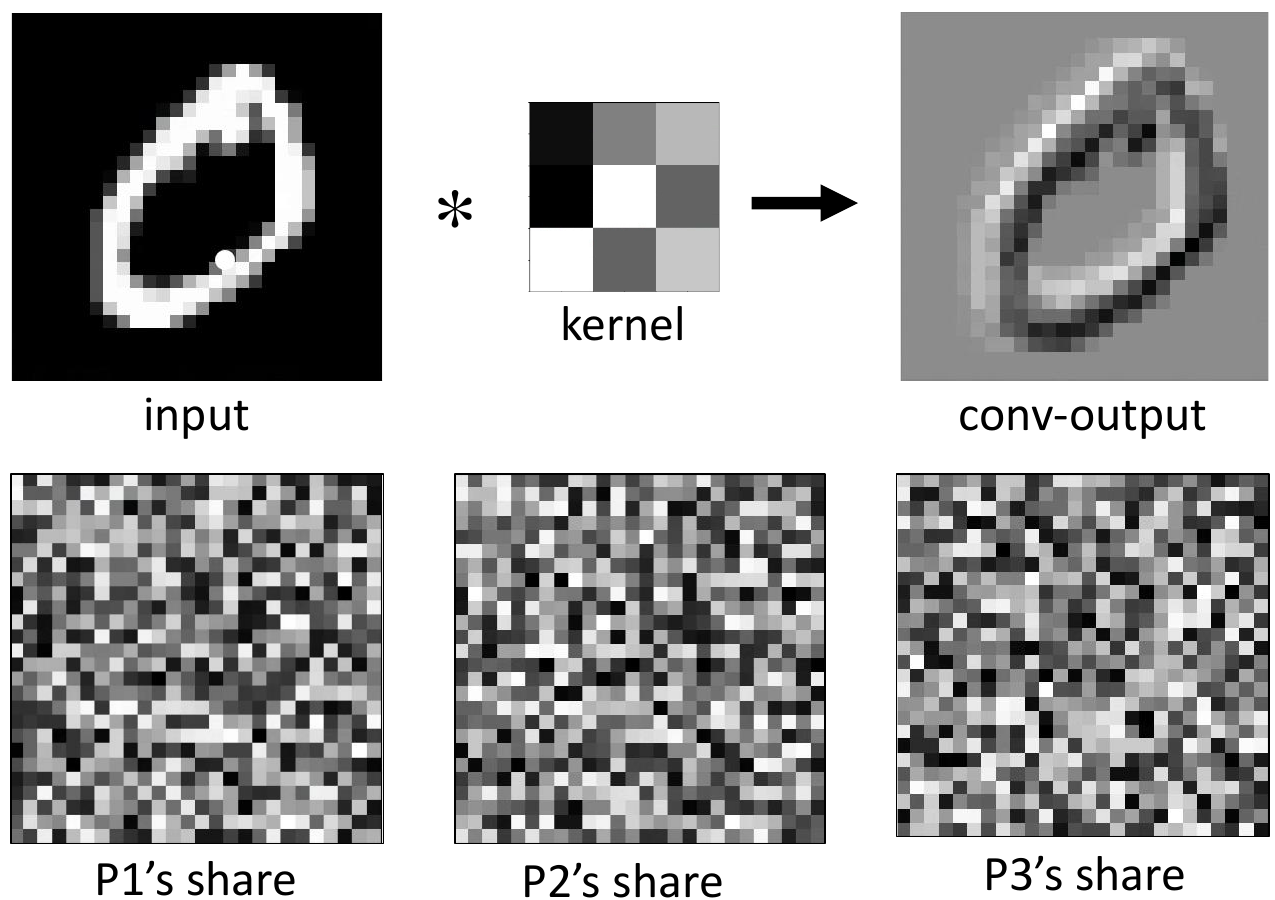}
    \caption{A tiny convolution example showing the result in plaintext (upper) and in Shamir's shares (lower), assuming 3PC. We randomly generate a kernel for 2-D convolution, yet the convolution output in plaintext can still disclose sufficient knowledge of the input image.}
    \label{fig:first_conv}
\end{figure}

As a straightforward demonstration in Fig.\ref{fig:first_conv}, the feature map (as the secret) after the convolution indeed reveals much knowledge of the input, while the corresponding Shamir's shares in each party show no meaningful pattern. If the output is leaked to the model vendor,
the user's data might be leaked by reverse engineering. On the other hand, if the user can have access to the convolution output, i.e., a malicious user has access to $x$ and $y$ in $y=Conv2D(x,w)$, model weights $w$ can be reasoned by continuously feeding in various inputs $x$ and obtain corresponding outputs $y$. Thus, it is necessary to keep the multiplication result secured as well, in order to ensure the privacy of both users' input data and the model's parameters.

\subsection*{B. Case Study for Truncation Error}
We provide an intuitive example here to demonstrate the error influence on additive sharing and SSS during truncation. Assume a finite filed $\mathbb{F}_{11}$ under 3PC, the truncation input as a secret is $a_0=5$, and the scaling factor $r=2$, so that the truncation in \textit{plain text} is $\lfloor a_0/r\rfloor=2$. We assume an additive sharing has shares $(9,9,9)$ so that $a_0=(9+9+9)\ \text{mod}\ 11=5$. During truncation, each party locally does the division and rounding, to get $(4,4,4)$ as shares; thus the reconstructed secret is $1$, which is still close to the original secret $a_0=2$. However, we assume the SSS has shares $(0,6,1)$ follow $(2,3)$-SSS scheme so that $\mathsf{REC}((0,6),2,2)=5$. If the truncation is also conducted locally, the shares become $(0,3,0)$. To run reconstruction $\mathsf{REC}((0,3),2,2)=8$, which is far from the correct truncation result.

\subsection*{C. Specified AvgPooling Operation}
In Sec.~\ref{sec:sss_nonlinear}, we discuss the approach of merging the division to the preceding SSS-Truncation operation. By this manipulation, small errors will be induced in the computation. In a normal truncation, $y=\left \lfloor x/r\right \rfloor$ is calculated; note that the masking is neglected in our discussion here since they are designed to be canceled perfectly. Focusing on only one kernel $y'$ with the kernel size $k_h\times k_w$, where $h,w$ is the row/column index, the ReLU+AvgPooling computes 
\begin{equation*}
    y_{avg}=\left[ \frac{1}{k_h k_w}\sum_{h=1}^{k_h}\sum_{w=1}^{k_w} \text{ReLU}(y'_{h,w}) \right]
\end{equation*}
for one kernel. However, our manipulation conducts
\begin{equation*}
    \widehat{y_{avg}}= \sum_{h=1}^{k_h}\sum_{w=1}^{k_w} \text{ReLU}\left (\left[\frac{y'_{h,w}}{k_h k_w}\right]\right ) 
\end{equation*}
where $[\cdot]$ means the rounding operation. To calculate the worst case of error, i.e., $|y_{avg}-\widehat{y_{avg}}|$, we assume each element in $y'$ is $y'_{h,w}=(g\cdot k_hk_w+ \lceil k_hk_w/2\rceil)$, where $g\in \mathbb{N}$ and $y'\in \mathbb{F}_p$; further all the elements in $y'$ are positive to retain the difference. Thus, $y_{avg}=g\cdot k_hk_w+ \lceil k_hk_w/2\rceil$ and $\widehat{y_{avg}}=(g+1)\cdot k_hk_w$. The difference between $y_{avg}$ and $\widehat{y_{avg}}$ is $k_hk_w-\lceil k_hk_w/2\rceil$, which is the magnitude of the error that our method will induce. Therefore, for a common $2\times2$ AvgPooling, the error for one kernel is at most $\pm2$.

\subsection*{D. Split Strategy}
\begin{figure}
    \centering
    \includegraphics[width=\linewidth]{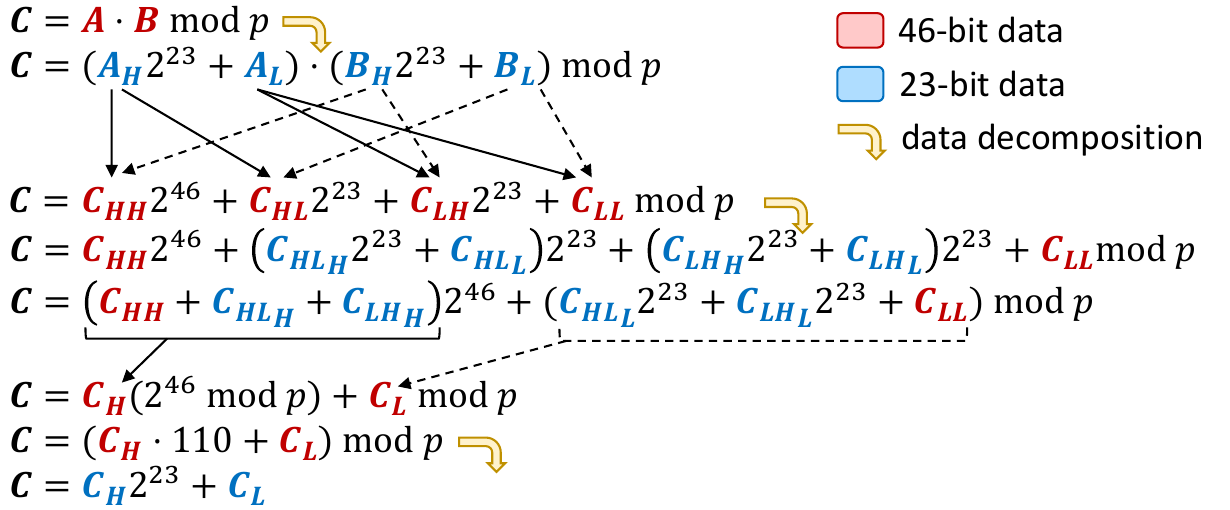}
    \caption{The multiplication on decomposed 46-bit data \textbf{A} and \textbf{B}. Only colored variables participate in computation, while the normal text (e.g., 2's exponents and mod) is given to prove computation correctness.}
    \label{fig:datadecomp}
\end{figure}

This is a supplementary explanation of the data decomposition strategy (see Sec.~\ref{sec:bit_width}) to perform 46-bit integer multiplication on $\mathbb{F}_p$ where $p$ is a 45-bit large prime number, as demonstrated in Fig.~\ref{fig:datadecomp}. Specifically, we calculate $\textbf{C}=\textbf{A}\cdot \textbf{B}\ \text{mod}\ p$ where $\textbf{A}, \textbf{B},$ and $\textbf{C}\in \mathbb{F}_p$. Since GPU only supports for 64-bit floating point, we split \textbf{A} and \textbf{B} into $\textbf{A}=\textbf{A}_H * 2^{23}+\textbf{A}_L$ and $\textbf{B}=\textbf{B}_H * 2^{23}+\textbf{B}_L$, so that $\textbf{A}_H, \textbf{A}_L, \textbf{B}_H$, and $\textbf{B}_L$ are 23-bit. 

We first calculate $\textbf{C}_{HH}=\textbf{A}_H\cdot \textbf{B}_H$, $\textbf{C}_{HL}=\textbf{A}_H\cdot \textbf{B}_L$, $\textbf{C}_{LH}=\textbf{A}_L\cdot \textbf{B}_H$, and $\textbf{C}_{LL}=\textbf{A}_L\cdot \textbf{B}_L$ in parallel, so that $\textbf{C}=\textbf{C}_{HH}2^{46}+\textbf{C}_{HL}2^{23}+\textbf{C}_{LH}2^{23}+\textbf{C}_{LL}$; note that the four multiplications here lead to 46-bit results. Intuitively, the modulo operation on $p$ can be directly applied to each term; however, for example, $\textbf{C}_{HL}2^{23}$ mod $p$ still exceeds the GPU 64-bit computation capability. On the other hand, $\textbf{C}_{HH}2^{46}$ mod $p$ can be well process, because $2^{46}$ mod $p=110$; thus $\textbf{C}_{HH}2^{46}($mod $p)\equiv\textbf{C}_{HH}110($mod $p)$. Thus, this observation derives a computing strategy that we can compute the modulo by splitting the summation above into two parts, larger than $2^{64}$ and less than $2^{64}$. Specifically, $\textbf{C}_{HL}$ and $\textbf{C}_{HL}$ are further decomposed as 23-bit splits, ${\textbf{C}_{HL}}_{H}$, ${\textbf{C}_{HL}}_{L}$, ${\textbf{C}_{LH}}_{H}$, and ${\textbf{C}_{LH}}_{L}$. The summation is reshaped as $(\textbf{C}_{HH}+{\textbf{C}_{HL}}_{H}+{\textbf{C}_{LH}}_{H})2^{46}+({\textbf{C}_{HL}}_{L}+{\textbf{C}_{LH}}_{L}+\textbf{C}_{LL})$. Therefore, \textbf{C} is calculated as $\textbf{C}_H2^{46}+\textbf{C}_L$ mod $p$. The modulo is conducted on each term. Note that the computing procedure above is always in the 64-bit GPU capability and runs each step in parallel.

\begin{figure}
    \centering
\subfigure{\centering
    \includegraphics[width=\linewidth]{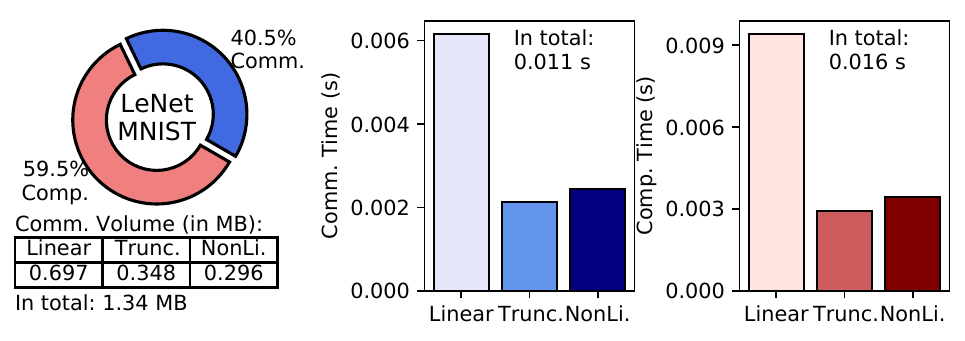}}
\subfigure{\centering
    \includegraphics[width=\linewidth]{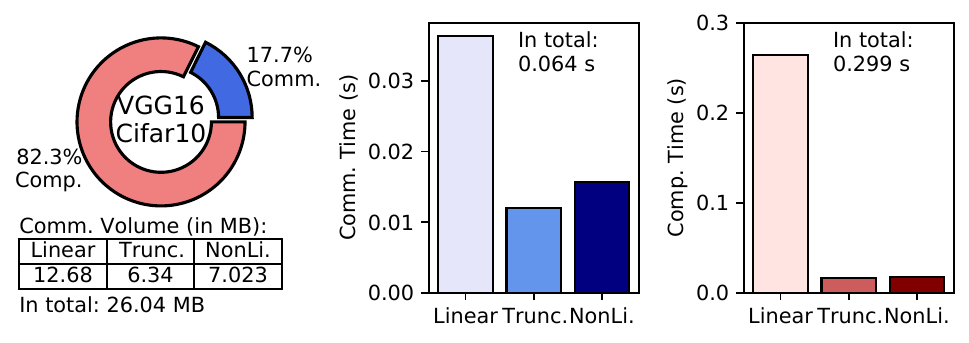}}
\subfigure{\centering
    \includegraphics[width=\linewidth]{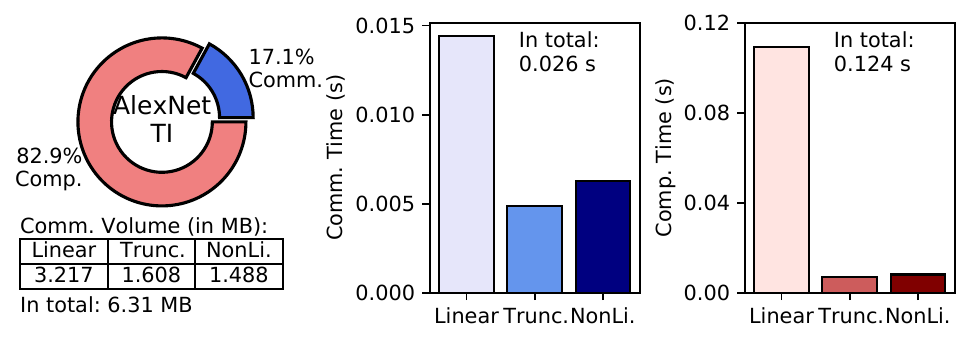}}
\subfigure{\centering
    \includegraphics[width=\linewidth]{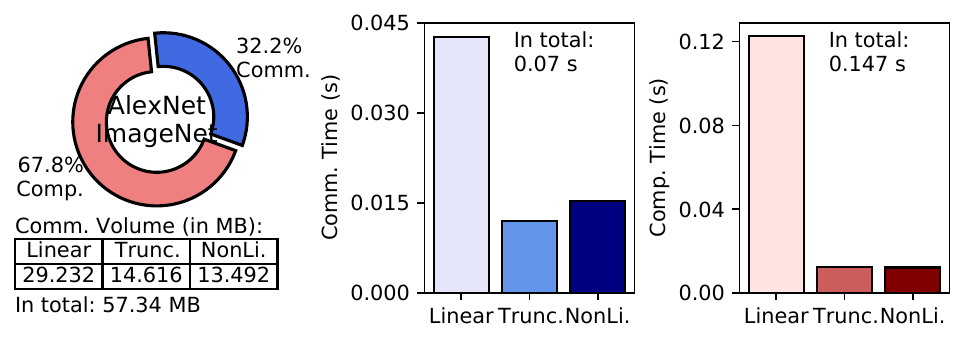}}
\subfigure{\centering
    \includegraphics[width=\linewidth]{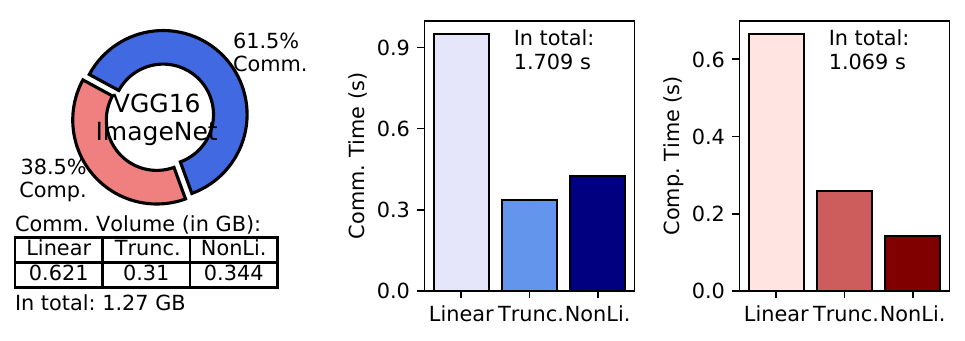}}
\subfigure{\centering
    \includegraphics[width=\linewidth]{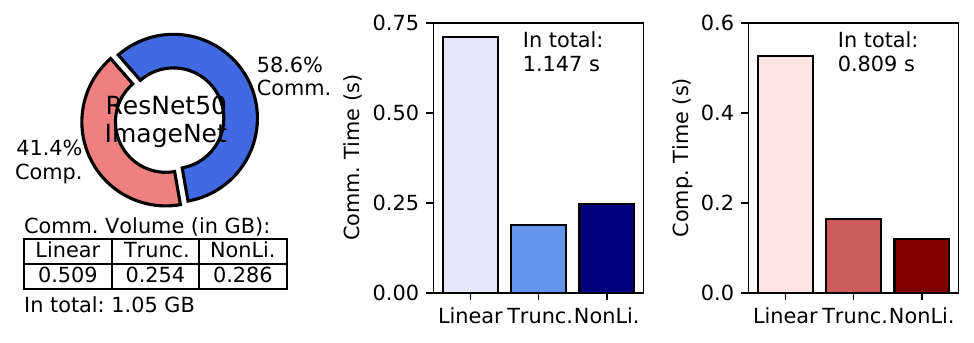}}
\subfigure{\centering
    \includegraphics[width=\linewidth]{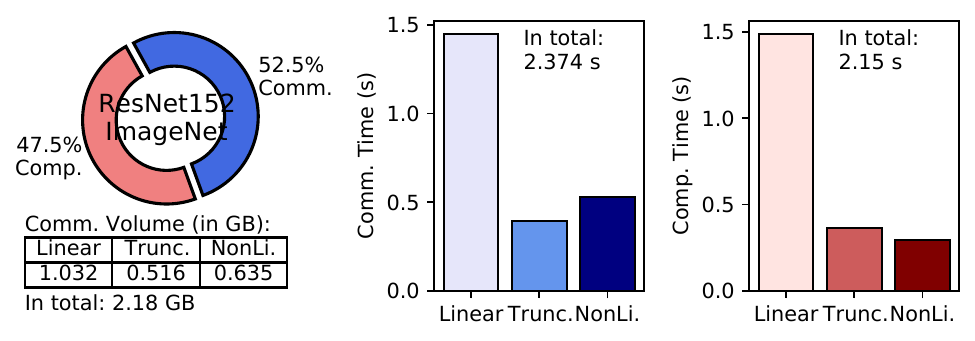}}
\caption{The performance decomposition of \ouralg framework on other tasks except which we selected in Fig.~\ref{fig:ablation_study}.}
    \label{fig:ablation_study_appendix}
\end{figure}

\subsection*{E. Ablation Study on Other Tasks}
We provide the performance decomposition analysis on all the other tasks in Fig.~\ref{fig:ablation_study_appendix}, except the selected tasks in Fig.~\ref{fig:ablation_study}. We can observe the same trend, that communication will take more time ratio during model inference on larger-scale tasks. For example, most of ImageNet tasks in Fig.~\ref{fig:ablation_study} and Fig.~\ref{fig:ablation_study_appendix} cost over half the time to conduct communication; yet, this ratio is only around $20\%$ on Cifar-10 and Tiny ImageNet classification. However, there is an exception that the communication time ratio of LeNet inference on MNIST is as large as $40\%$. This is because the model scale of LeNet is so lightweight that GPU can substantially optimize the computation; thus the computation time is quite low, making communication have a large time ratio. This can be evidenced by the fact that, comparing the computation time in Fig.~\ref{fig:ablation_study} and Fig.~\ref{fig:ablation_study_appendix} with the plaintext execution time in Tab.~\ref{tab:performance_comparison}, LeNet (MNIST) only increases the computation time $\times 23$ (from $0.7$ms to $16$ms). However, the computation time increment is much larger on other tasks, such as $\times 76$ on AlexNet (Cifar-10), $\times 149$ on VGG-16 (Tiny ImangeNet), and $\times 72$ on ResNet-101 (ImageNet).

\end{document}